# Identification of materials with strong magneto-structural coupling using computational high-throughput screening

Luis Casillas-Trujillo, Rickard Armiento, Björn Alling
Department of Physics, Chemistry and Biology (IFM), Linköping University, 58183 Linköping, Sweden.

## Abstract

Important phenomena such as magnetostriction, magnetocaloric, and magnetoelectric effects arise from, or could be enhanced by, the coupling of magnetic and structural degrees of freedom. The coupling of spin and lattice also influence transport and structural properties in magnetic materials in particular around phase transitions. In this paper we propose a method for screening materials for a strong magneto-structural coupling by assessing the effect of the local magnetic configuration on the atomic forces using density functional theory (DFT). We have employed the disordered local moment approach in a supercell formulation to probe different magnetic local configurations and their forces and performed a high-throughput search on binary and ternary compounds available in the Crystallographic Open Database. We identify a list of materials with a strong spin-lattice coupling out of which several are already known to display magneto-lattice coupling-phenomena like $Fe_3O_4$ and CrN. Others, such as $Mn_2CrO_4$ and $CaFe_7O_{11}$ have been less studied and are yet to reveal their potentials in experiments and applications.

## 1. Introduction

Magnetic materials are used in a wide range of applications including novel and attractive technologies where materials improvements are desired. Yet, they also represent a great challenge in fundamental physics and computational simulations slowing down the phase of development in areas such as magnetic refrigeration, permanent magnets and steels. In particular, in theoretical first-principles research these issues manifest themselves in materials-specific computational and numerical challenges, which is reflected in the relatively low number of theoretical high-throughput magnetic material studies. Even standard low-throughput magnetic calculations are challenging, often leading to simplifying assumptions, such as neglecting magnetic excitations and disorder and even approximating the paramagnetic state as non-magnetic [1,2]. When the complexity of magnetism is indeed investigated, lattice degrees of freedom, like vibrations or lattice relaxations are often neglected [3]. However, many of the technologically attractive phenomena displayed by magnetic materials originate from the coupling between magnetic, electronic, and structural degrees of freedom. Magnetostriction, the magnetocaloric effect (MCE), and the magnetoelectric effect are already being used in technological applications and play a fundamental role in the development of new technologies [4].

The coupling phenomenon between the magnetic field and the strain field of magnetostrictive materials make them widely used in sensors and actuators [5]; applications range over, e.g., load cells, torque meters, pressure gauges, tensile stress sensors, thin-film thickness sensors, and dynamic stress sensors [6]. Taking advantage of this phenomena can also improve or revolutionize well-stablished technologies. Magnetostrictive transducers were recently developed with the unique characteristic that they do not require direct wiring between the magnetostrictive material and the sensing/actuating solenoids [7].

The magnetoelectric effect has applications in memory devices, magnetic field sensors, and spintronics [8-11] by providing an efficient control of magnetism by an electric field. Switching the orientation of magnetization using an electric field has a substantially lower energy requirement [12]. The ability to modify the magnetization using a low magnitude electric field (and vice versa) is of great interest for speeding up the processing rate of memory devices, and also for the design of very sensitive stress sensors [13].

Magnetocaloric materials have a strong coupling between the crystallographic structure and magnetism. A magnetic field can induce a simultaneous change of magnetic and lattice entropies [14]. In the well-known case of $Gd_5Si_2Ge_2$, a magnetic-field-induced transformation from the paramagnetic monoclinic phase to the ferromagnetic orthorhombic phase, the entropy change due to structural transition contributes to more than half of the total entropy change [14]. Magnetic refrigeration, based on the MCE, has been proposed as an energy efficient and environmentally friendly alternative to vapor-compression refrigeration. [15]. Developments in magnetocaloric materials in recent years have led to the first prototype systems [16], yet material limitations, like component weight and cost, are still an obstacle to allow large scale commercial distribution. Besides the application of the MCE to magnetic refrigeration equipment, magnetic heat pumps, and magnetocaloric energy conversion machines, the MCE has shown great promise for medical applications, targeted drug delivery and, the destruction of cancer cells [17].

Besides these phenomena, the coupling of magnetic and structural degrees of freedom impacts material properties such as thermal conductivity. The anomalous thermal conductivity exhibited by magnetic semiconductors has been recently attributed to the shortening of the lifetime of acoustic phonons due to spin-phonon coupling [18]. The thermal conductivity of CrN displays an extremely different temperature dependence below and above the magnetic transition temperature, remaining almost constant above the transition temperature. The dynamics of the spin system severely impact lattice vibrations, reducing the phonon lifetime of the acoustic modes above the transition temperature significantly. This behavior has also been observed in one of the top candidates found in the high-throughput study, $Fe_3O_4$, which displays a strong change in the behavior of thermal conductivity and thermal diffusivity close to the transition temperature [19]. The relationship between structure and magnetism is also present in crystallographic defects, where the change in environment carries a concomitant change in magnetic moment as in the case for screw dislocation in the paramagnetic state [20].

High-throughput methods have proven to be a powerful tool in materials science, allowing the discovery and design of new materials [21-24]. Furthermore, theoretical high-throughput calculations can efficiently screen a large quantity of materials without experimental synthesis [25], representing savings in time and money by pointing to the most promising candidates that can be further examined by experimental investigations. While high-throughput methods sacrifice accuracy for automated analysis of a large number of compounds, they can produce more candidates than by high-accuracy low-throughput studies. The vast amount of material data already available in databases offers the possibility to efficiently screen a large quantity of materials. As such, it enables probing materials for different target properties. Hence, it is possible to find new applications for well-known materials, for example, the case of $UO_2$ which is famously known for its use as nuclear fuel, but is also the hardest piezomagnet known [26].

The paramount importance of magnetic materials has motivated several experimental high-throughput searches to find better performing magnets [25,27] and the development of machine learning approaches to improve high-throughput searches of rare-earth free permanent magnets [28]. As mentioned before, despite all the fascinating properties displayed by magnetic materials, there have been few theoretical high-throughput calculations. Recently Horton et al. [29] performed a high-throughput study to assess the magnetic order of the ground state in inorganic materials, and Bocarsly *et al*. [30] screened magnetocaloric compounds. In this work we conduct a high-throughput search to find materials with strong magneto-structural coupling using density functional theory (DFT) calculations within the High-throughput Toolkit (*httk*) framework [31]. To assess magneto-structural coupling we suggest the following descriptor: The effect that the different local magnetic environment in a disordered magnetic structure has on the force acting on each atom in the structure. This is a different effect compared to the magneto-structural coupling happening through the spin-orbit coupling. The relativistic spin-orbit coupling is weak on an atomic scale but important in deciding the global magnetization direction with respect to the crystal structure in ordered magnetic states, manifested in the magnetic anisotropy. By using here the magnetically disordered state we probe the atomic-scale, electronic origin, of spin-lattice coupling in a model of the paramagnetic state of direct relevance for the magnetocaloric effect and thermal conductivity at high temperature. But simultaneously, the use of the symmetry breaking of the disorder act as a path to discover spin-lattice coupling, of possible importance also for ordered magnetic states, that might be hidden in calculations if they are performed on highly symmetric ground-states.

## 2. Methods

### *2.1 Disordered local moment*

The simulation of the paramagnetic state by first-principles calculations is challenging and, in some cases, has been approximated by the non-magnetic state. The interatomic bond strengths can differ greatly between magnetic and non-magnetic states [32], leading to incorrect results [1,2]. In order to properly describe the paramagnetic state, one can employ the disordered local moment approach (DLM)[3] as implemented in a supercell framework [33,34]. The real magnetization density in many magnetic materials can be described in terms of magnetic moments localized close to the atoms. The PM case can then be viewed as a disordered distribution of such local moments with no long-range order, and sufficiently above the critical temperature approximated with the fully disordered state lacking also short-range-order (SRO), i.e:

$$DLM\colon \frac{1}{N}\sum_{i,j\in\alpha} \mathbf{e}_i \cdot \mathbf{e}_j = 0, \forall\, \alpha. \quad (1)$$

where *N* is the number of magnetic pairs in the summation, *i* and *j* denotes the atoms in question belonging to a specific coordination shell $\alpha$. $\mathbf{e}_i$ and $\mathbf{e}_j$ are the unit vectors in the direction that maximizes the magnetization density around atom *i* or *j*, i.e. the direction of the atomic magnetic moment on atom *i* or *j*.

The DLM result can be obtained with a supercell constructed using the special quasirandom structures (SQS) approach [35] that fulfills Eq. 1 by construction for several closest coordination shells, or by sampling and averaging over a large number of randomly chosen supercell configurations [34]. The latter is more practical in the present case as the SQS approach demands the design of a separate SQS for each of the hundreds of crystal structures studied here. For the high-throughput screening, we have chosen to sample a few random configurations since this represents saving in time and computer resources, and the descriptor is not evaluated with a need for full convergency with respect to magnetic SRO-statistics. The DLM approach allows formulating a consistent electronic-structure based thermodynamic theory, that accounts for the interplay between the configurational and magnetic degrees of freedom [36]. The DLM approach can be employed with either collinear or non-collinear magnetic moments [34]. In general, non-collinear calculations give a smaller magnitude for the forces than collinear calculations [37], yet the choice of approach has no effect on the force based hierarchy among the probed compounds. Non-collinear calculations are significantly more computationally expensive than collinear calculations, and therefore we have chosen to perform the calculations with collinear magnetic moments.

### *2.2 Descriptor: Interatomic forces induced by disordered magnetism*

Filipetti *et al.* [38] proposed the concept of magnetic stress, defined as the stress associated with the change in spin ordering. This concept establishes a connection between magnetic ordering and structural properties. Bocarsly *et al.* [39] suggested that magneto-structural coupling has an important role behind the magnetocaloric behavior, and employed the magnetic deformation [30], a DFT based descriptor defined as the degree of lattice deformation between non-magnetic and ferromagnetic unit cells, to assess the magnetocaloric effect in MnB and FeB. They found that while MnB is isostructural to FeB and has similar magnetic properties, the former possesses a larger magnetic deformation and has a larger magnetocaloric effect, whereas the latter displayed little magneto-structural coupling.

Yet these approaches do not capture the effect of magnetic disorder and local magnetic environment effect. Important structural transitions occur at the critical temperature where the system goes from paramagnetic to some ordered magnetic state, and as such one should investigate the paramagnetic state. Therefore, we use the force induced on each atom by the local magnetic configuration in the paramagnetic state as a descriptor to find materials with strong magneto-structural coupling. We suggest that if the change in the local magnetic configurations produces a large change in the force value, it is an indicator of strong magneto-structural coupling. With this approach we isolate the effect of magnetism on interatomic forces, and the relief of these forces may come as structural transformations or manifest itself as a dynamical coupling between lattice and spin dynamics [18]. In principle, this approach can be employed between different ordered magnetic states. However, due to the symmetry in ordered cases, some of the spin-lattice coupling induced forces can be hidden. Employing the disordered magnetic state gives direct insight into the physics around the order-disorder transition, critical in many applications of magnetic materials.

To illustrate the dependance of the force acting on each atom on the magnetic environment we use Chromium nitride (CrN), bcc-Fe and bcc-$Fe_{0.5}Cr_{0.5}$. CrN is a prototype material with strong magneto-structural interactions. At room temperature it has a rock salt structure and paramagnetic state. Below its Neel temperature (Tn=273K) it undergoes a structural and magnetic transition to antiferromagnetic (AFM) ordering and to an orthorhombic structure with space group Pnma [40-43]. In Figure 1 we compare and contrast the behavior of CrN, bcc-Fe and bcc-$Fe_{0.5}Cr_{0.5}$ random alloy, where we plot the magnitude of the net force acting on each atom sitting on ideal lattice positions but with a disordered magnetic configuration. In CrN, it can be seen that there is a large oscillation in the force values of Cr atoms, indicating a strong dependence on the local magnetic configuration, having a range of ~1 eV/Å and an average value of 0.56 eV/Å for the Cr atoms, while the force on the nonmagnetic N atoms is considerably lower with an average of 0.20 eV/Å. For bcc Fe and FeCr, the force on Fe atoms has a similar value, with 0.25 eV/Å for bcc Fe and 0.19 eV/Å for FeCr, and an average value of 0.09 eV/Å for Cr atoms in FeCr. Thus, Cr atoms in CrN present a distinctively larger average value and force range. These force values highlight that the descriptor captures the combined effect of the symmetry-broken disordered magnetic state with different magnetic local environments of each atom and the general bond-strength of the material which is high for the ceramic CrN. Hence, the descriptor is suitable for searching for materials with strong spin-lattice coupling for use as multiferroic, magnetocaloric, and magnetostrictive materials.

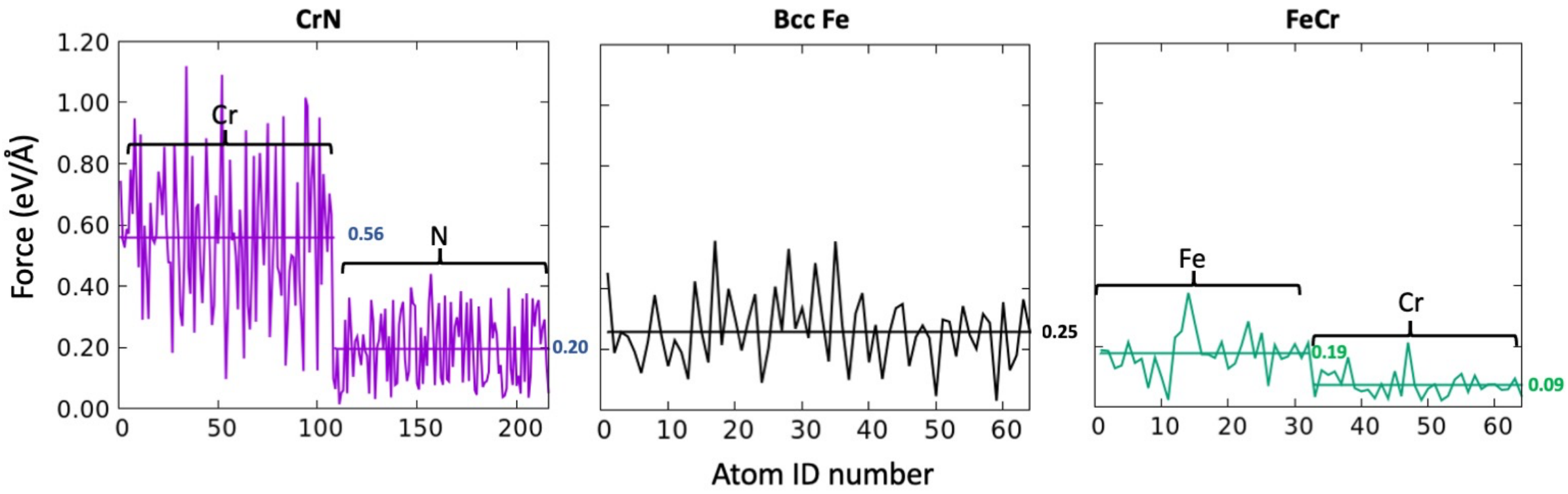

Figure 1. Comparison of forces in the paramagnetic state between CrN, bcc Fe, and bcc $Fe_{0.5}Cr_{0.5}$ alloy.

*2.3 Computational details*

The set of compounds to perform the high-throughput calculations are obtained from the Crystallographic Open Database (COD) [44]. We have retrieved binary and ternary compounds that contain the magnetic elements Ni, Cr, Co, Fe, and Mn. From this set we select compounds with at least 20% magnetic elements per formula unit. To have enough atoms for magnetic disorder, we require supercells with at least 10 magnetic atoms. Thus, if the original structure is small, a supercell of at least 10 magnetic atoms is created by multiplying the unit cell in x, y, and z until the minimum size is attained. The high-throughput calculations are performed using the high-throughput toolkit (*httk*) which interfaces with the Vienna *ab-initio* package (VASP) [45,46] to perform the *ab initio* collinear spin-polarized calculations with the Perdew–Burke–Ernzenhof generalized gradient approximation [47] to model the exchange-correlation effects and the projector augmented wave basis [48]. The structure of the selected compounds was first relaxed using a ferromagnetic configuration to assess the structural differences between the structure reported in the database and the relaxed structures from first-principles calculations. The energy cutoff is set to 600 eV for all compounds and we employed Monkhorst-Pack *k*-point grid with 20 *k*-points/Å density for all supercells. We do not go beyond standard DFT calculations, and therefore we do not employ DFT+U corrections and excluded structures with chemical disorder, e.g. atomic sties with partial occupancies.

Once we obtain the ferromagnetically relaxed structure, we impose a randomly assigned DLM magnetic configuration with local magnetic moments initiated with a 3 Bohr magnetons magnitude. We then perform a static calculation to obtain the forces on the atoms. We have set a threshold of 0.3 eV/Å to decide if the material displays large forces. This value was set during our preliminary calculations which included 100 compounds chosen by hand. If any of the individual forces in the material passes the threshold, we perform a second calculation with a different DLM configuration, with the goal of comparing the difference in forces caused by different disordered configurations that differ only by local magnetic configurations while having the same lack of long range order and magnetic symmetry. Otherwise, we deem the material as having weak magneto-lattice coupling and it is screened out. This screening procedure is schematically shown in Figure 2. In these collinear calculations, the sign of the magnetic moments is not constrained. After the calculations, we consider as magnetic species the elements that possess an average magnetic moment above 0.05 Bohr magnetons. If the final magnetic moment of a considerable fraction of these magnetic atoms changes sign, it may be an indication of a strong magnetic ordering tendency, could reduce the difference in local magnetic environments, and cast some questions if its forces can be compared to the non-flipped, fully disordered cases. Obviously, if 50% of the moments would flip, any ground state ordered magnetic configuration could have appeared. But for our randomly assigned starting configuration, even with less flips, disorder might be drastically reduced. For this reason, we focus here on the materials with less than 20% spin-flips. However, all materials that pass the force-criteria are presented in the appendix [49] together with the fraction of spin-flips.

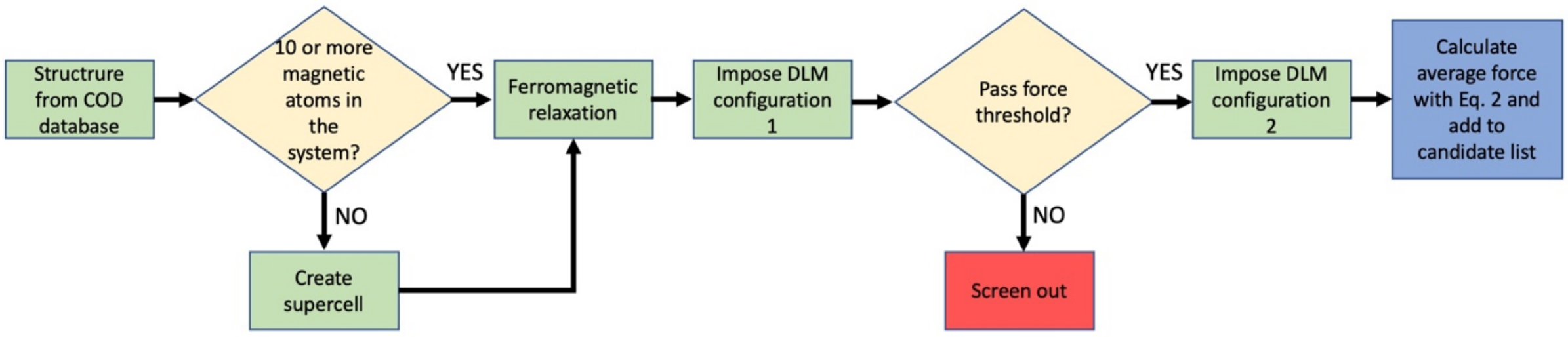

Figure 2. Schematic representation of the screening process.

## 3. Results

### *3.1 Single elements*

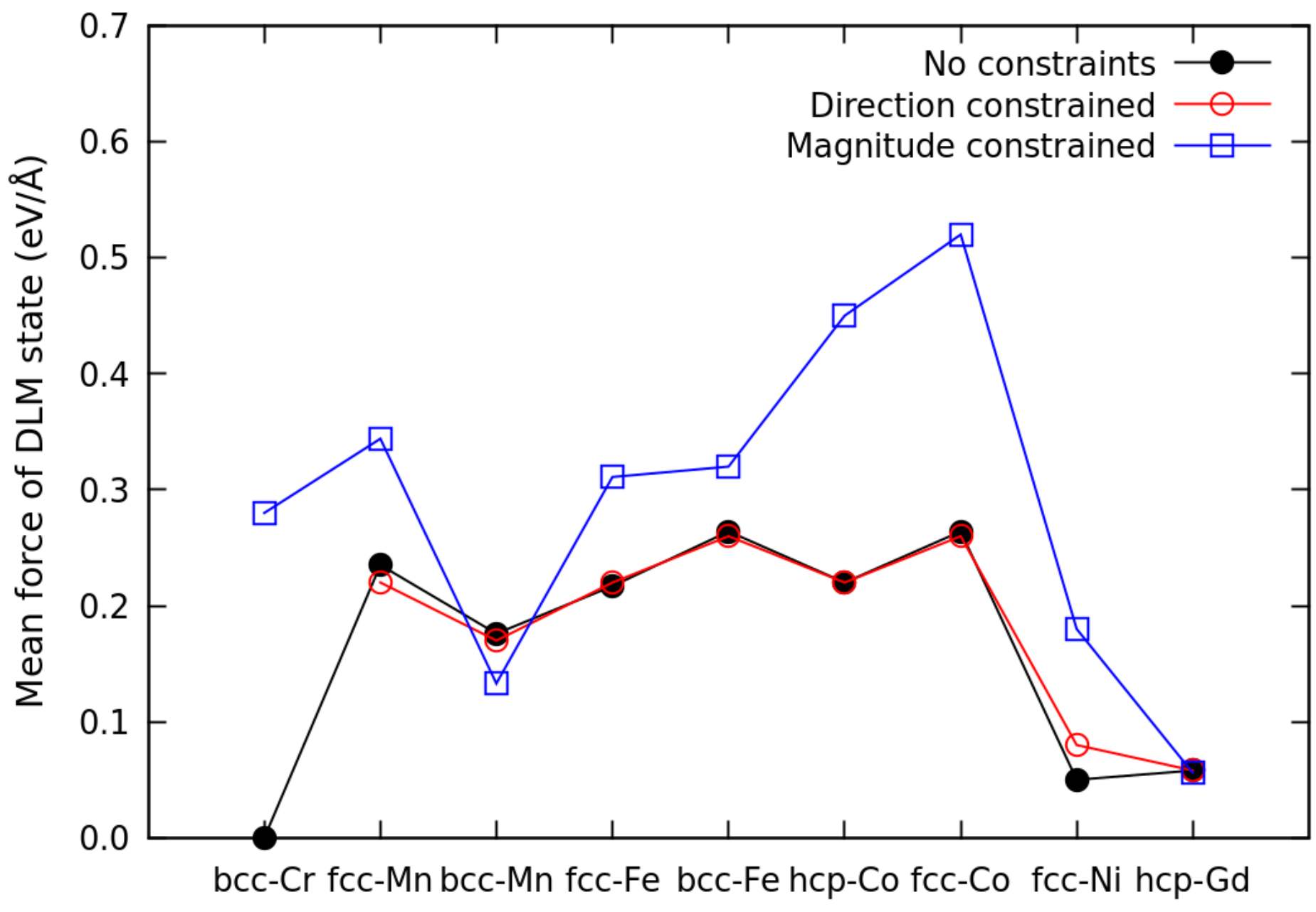

Figure 3. Mean force for pure magnetic elements with unconstrained, direction constrained, and fully constrained magnetic moments.

First, we investigate the magnetically induced forces for single magnetic elements. We have considered bcc-Cr, fcc-Mn, bcc-Mn, fcc-Fe, bcc-Fe, hcp-Co, fcc-Co, fcc-Ni and, hcp-Gd, and perform calculations with unconstrained magnetic moments, magnetic moments constrained in the direction up/down, and with constraints in both direction and magnitude. Henceforth, we will refer to the latter as fully constrained calculations.

Focusing first on the unconstrained calculations we see that Mn, Fe, and Co all display magnetically induced forces in the order of 0.2-0.3 eV/Å regardless of considered crystal structure. Bcc-Fe and fcc-Co share the first place in the unconstrained calculations with forces of 0.26 eV/Å. Fcc-Ni and hcp-Gd have low values of magnetic disorder induced force, 0.05 eV/Å and 0.06 eV/Å respectively. The moments in bcc-Cr collapse to zero in these unconstrained calculations and thus there are zero induced forces in that case.

In the unconstrained calculations a few flipping of the magnetic moments occur. By constraining the direction of the magnetic moments, we can evaluate the effect of magnetic moment flipping on the average force. There is very little difference in the force values between the unconstrained and direction constrained magnetic moments as shown in Figure 3, suggesting that a few magnetic moment flipping has not a pronounced effect on the forces. For the fully constrained calculations we use the magnitude of the ferromagnetic moments. In this case the results differ from the other two cases, presenting in general, a larger value for the average force, with a more pronounced effect on itinerant magnetic systems. The exception is bcc-Mn, which displays a lower average force value for the fully constrained case. The lower force value is due to that the magnetic moment in the disordered state is larger than in the FM case. Likewise, the large increase in force in the fcc-Co case is due to the considerably larger FM magnetic moment magnitude, having a value of 1.7$\mu_B$ while in the disordered case the average magnetic moment is 1.07$\mu_B$. We have used constrained calculations as a mean to assess the effect of magnetic moment magnitude and direction. Based on the findings for these elements we note that unless there is a dramatic collapse of the magnetic moments in the DLM or the FM state, the increased control of the magnetic moments by external constrains do not change the relative hierarchy of magnetic disorder induced forces. Furthermore, it is artificial to demand that the moment size should be identical between different magnetic states, e.g., FM and PM in reality. We thus performed the high-throughput search using unconstrained magnetic moments but monitor the possible appearance of moment collapses and spin flips.

### *3.2. Prototype materials*

Following the elemental magnetic metals, we have tested the descriptor by probing archetypical materials for different magnetic properties of interest. FeRh is known for its magnetocaloric effect, $Fe_{64}Ni_{36}$ for its invar effect, and $BiFeO_3$ for its magnetoelectric effect. In Figure 4 we show the forces on each atom of the supercell, and the initial and final magnetic moments of each atom. In the case of FeRh, Fe atoms are subjected to larger forces than Rh atoms. Fe atoms possess an average magnetic moment close to $3\mu_B$. The magnetic moment on Rh atoms is small with a value close to $0.22\mu_B$. This compound suffered from many spin flips on the Rh atoms. Even if the magnetic moments are small, they prefer to be aligned in the same direction as the net magnetic moment of the cell, and we observe a similar behavior for $Fe_{64}Ni_{36}$. The force on Fe atoms was larger than Ni, with Fe atoms possessing an average magnetic moment of $2.3\mu_B$ and $0.3\mu_B$ for Ni atoms. The ordering tendency in Ni atoms is stronger than in the Rh case for FeRh. Finally, $BiFeO_3$ showed very large forces acting on Fe atoms, also large forces acting on the O atoms, while almost no force acting on Bi atoms. In this case O and Bi atoms has no magnetic moments while Fe atoms displayed a small increase from the initial $3\mu_B$ magnetic moment. All of these prototype compounds fulfilled the descriptor conditions, with $BiFeO_3$ presenting the largest average force among this model compounds, and larger than all the elements in the unconstrained results.

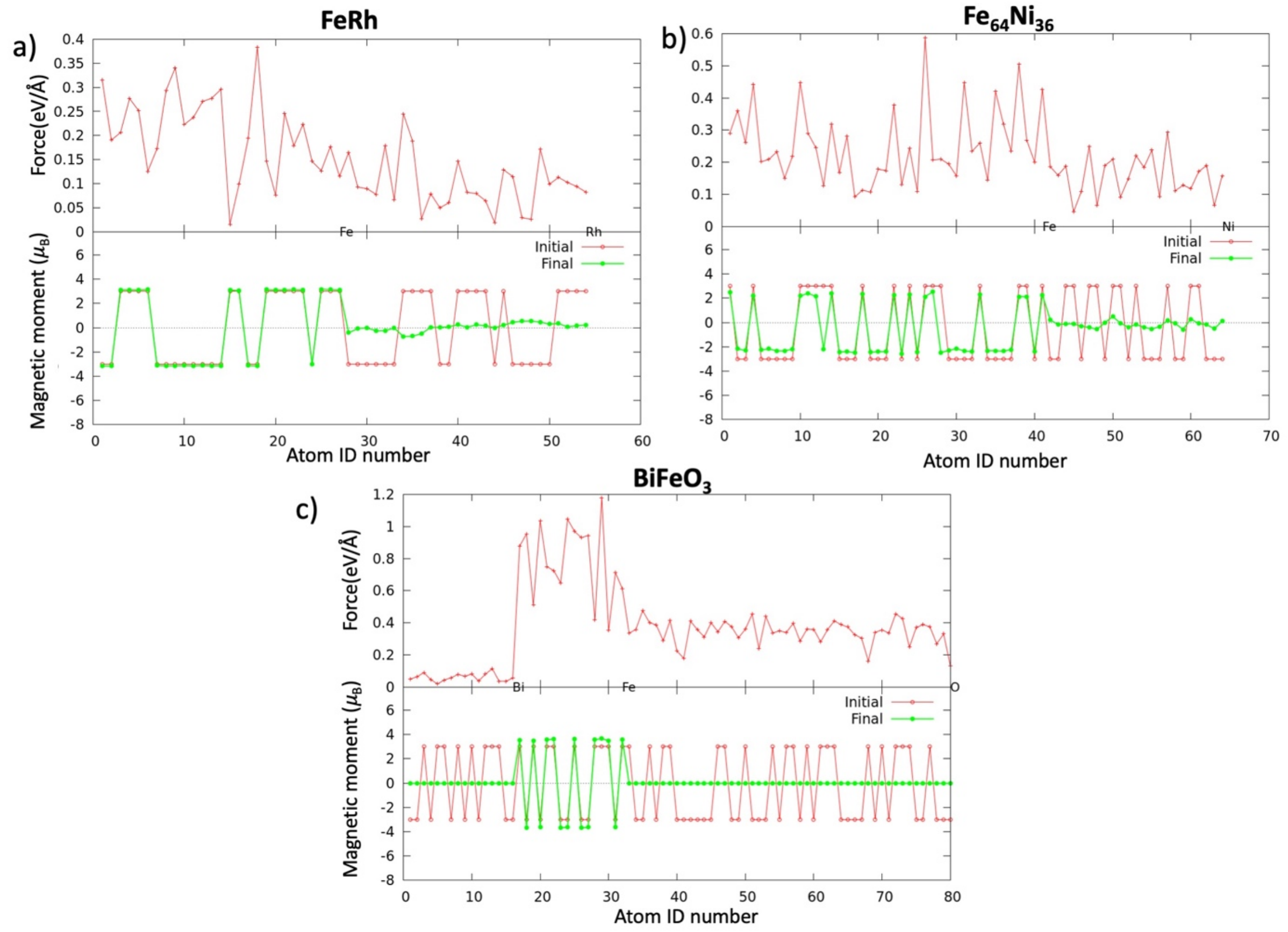

Figure 4. Force and magnetic moments acting on each atom in a) FeRh, b) $Fe_{64}Ni_{36}$, c) $BiFeO_3$.

*3.3 High-Throughput search*

We analyzed 1185 compounds from the COD database containing at least one Fe, Ni, Cr, Co, or Mn atomic species, with a 70% rate of successfully converged DFT calculations. We define a success as a converged DFT calculation. The main reason for non-convergence was due to the instability created by assuming a ferromagnetic configuration for the relaxation calculations. Our scheme could be improved by incorporating the methodology of Horton *et al.* [29] and first search for, and then use the obtained magnetic ground state as the starting point. From the set of successful calculations, we have 195 compounds containing Fe, 246 with Ni, 83 with Cr, 128 with Co and 206 with Mn, with 802 compounds containing only one magnetic element, and 28 with 2 magnetic elements. Applying our descriptor to these successfully converged calculations yielded 330 compounds of which 202 presented less than 20% spin flips.

We have ranked the candidate materials using the average force per atom given in equation 2, where the average is over all atomic species averaged over the two DLM configurations. In this setting a higher average force gets a higher ranking in the results list. We decided to include all atomic species, since in some cases, the highest forces in the structure are acting on the “non-magnetic” species, e.g., $CrMn_2O_4$ where the strongest forces appear on the O atoms. We consider an element as magnetic if the magnitude of its average magnetic moment is higher than $0.05\mu_B$ after relaxation. In Figure 5 we show the distribution of the average force, while in Figure 6 we plot the distribution of the force on magnetic elements, where it can be observed, that as a general trend, Fe atoms more often experience strong magnetic disorder-induced forces. In the top results list of Table 1, we have included compounds that presented 20% or less spin-flips. Even if our results for the elements above, and detailed constrained moments investigations of 20 additional materials, indicate that magnetic moment flipping has no strong impact on the average force, we have excluded compounds that have a large number of flips in the detailed discussion below. This is as their final magnetic state is less controlled and a material-specific detailed investigation is needed for a quantitative assessment of their magneto-structural coupling in the disordered magnetic state. Nevertheless, those that display a strong average force are included with a warning remark in the full list of compounds in the appendix [49], as a service for continued future studies.

$$\bar{F} = \frac{1}{N}\sum_{i}^{N}\left|\vec{F_i}\right| \quad (2)$$

We have also compared the pressure between the ferromagnetic and the DLM cases. Since we relax the ferromagnetic structure, the pressure for this phase is very close to zero. The magnetic moment magnitude changes between the ferromagnetic and the DLM states but not dramatically, therefore the pressure difference is not caused by the collapse of the magnetic moments in the DLM into a non-magnetic state. In figure 7a, we show a histogram ranking the candidate compounds according to their magnetic disorder induced pressure difference $\Delta P = P_{DLM} - P_{FM}$. A large pressure difference would be indicative of a driving force for a large coupling between magnetic order and/or excitations and volume change. In Figure 7a, we have included as a reference the values of bcc Fe and $Fe_{64}Ni_{36}$ prototype materials. $Fe_{64}Ni_{36}$ displays a negative pressure with respect to the ferromagnetic case for paramagnetic high temperature limit case. This is in line with the invar behavior of $Fe_{64}Ni_{36}$ where some, yet not fully established, magnetic excitation away from the FM state towards the disordered state give rise to a decrease in volume that almost exactly compensate for the normal thermal expansion due to anharmonic vibrations. Thus, the other magnetic materials observed to have a strong negative DLM-pressure can be candidates for invar-like behavior. There is a relationship between the increase/decrease of the magnetic moment in the DLM state with respect to the ferromagnetic moment and the pressure of the system in the DLM state, as can be seen in in Figure 7b. Although most compounds follow this trend, there are some points that greatly deviate. This is a consequence of the approximation of using the ferromagnetic state as the magnetic state in the relaxation. As mentioned before, this can be improved, to a computational price, by using the actual magnetic ground state with the method proposed by Horton *et al.* [29]. This approximation does not affect the hierarchy of the magneto-structural coupling list, since the results for this list are obtaining by examining the forces due to different DLM configurations without taking the magnetic ground state as a reference point.

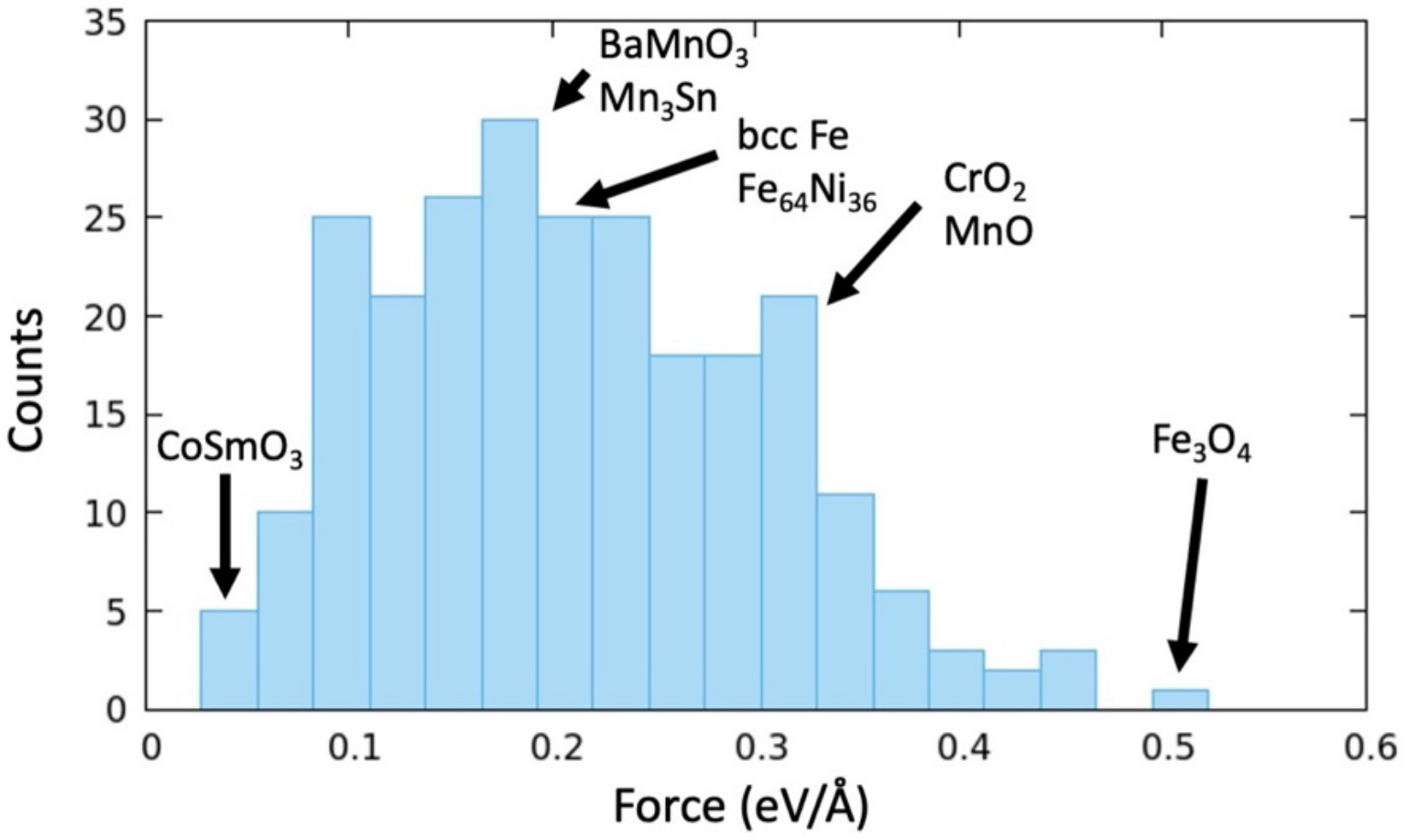

Figure 5. Distribution of the average force acting on each atom for the all the compounds in the candidate list.

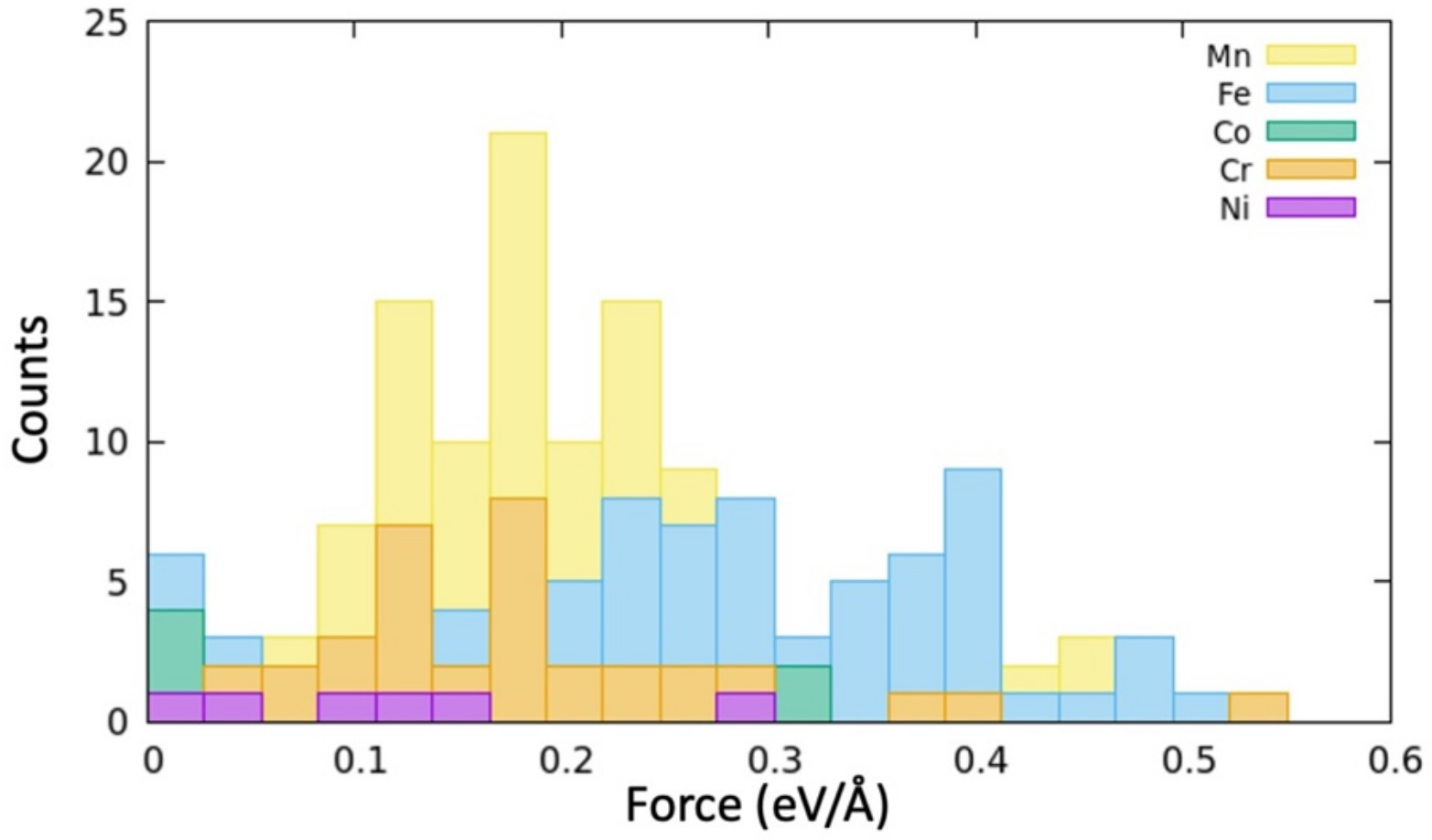

Figure6. Distribution of forces acting on the magnetic elements for the all the compounds in the candidate list.

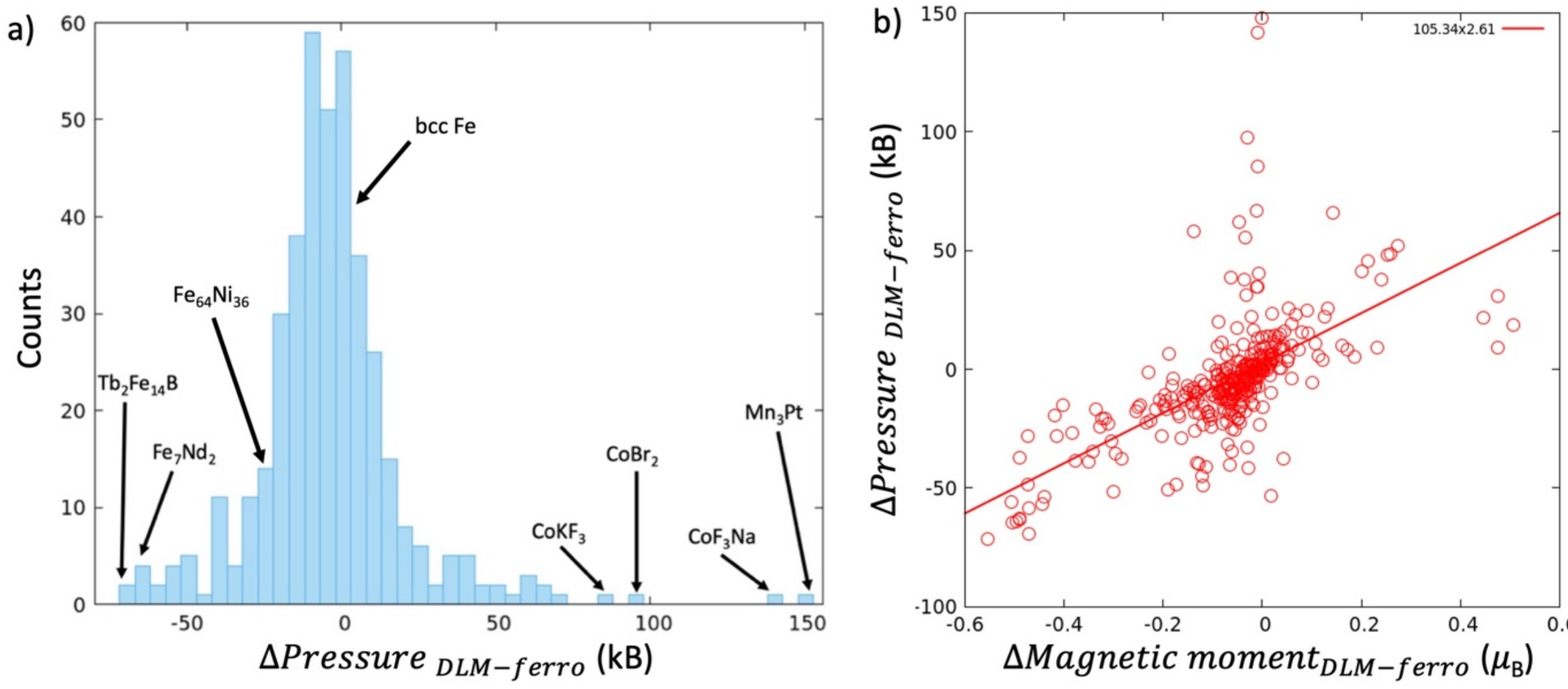

Figure 7. a) Distribution of the difference in pressure between the DLM and FM states for candidate compounds. b) Pressure difference versus magnetic moment difference between DLM and FM states.

The high-throughput search results yield materials that have been reported for magneto-structural coupling, and display technological relevant magnetic effects, including compounds that exhibit magnetostriction, magnetoelectricity, piezomagnetism, and superconductivity. The identification of these known compounds gives us confidence that the proposed descriptor is able to identify materials with magneto-structural coupling. In Table 1 we present the top 20 candidates from our high-throughput search. In this list we have only included compounds that present 20% or less spin-flips of the local magnetic moments. Not surprisingly, our prototype material CrN is highly ranked in the list and is included among the 20 highest ranked candidate materials. Among the pure elements the highest ranking are bcc-Fe and fcc-Co tied at place 93. Fcc-Mn, hcp-Co and fcc-Fe occupy the 117, 135 and, 138 positions respectively. Bcc-Mn ranks lower in the list at the 196th place, while fcc-Ni, hcp-Gd, and bcc-Cr are at the bottom of the list. The full list is provided in the appendix [49], where we also as a service to chemists present the top candidates for the oxides, nitrides and halogen compounds included in the search.

In contrast to the extensively researched CrN and $Fe_3O_4$, compounds among the top candidates like $CrMn_2O_4$ and $Ca_2Fe_7O_{11}$ have less comprehensive literature, with investigations mainly devoted to characterization [50,51], opening the opportunity for further investigations of their magneto-structural coupling related properties.

Table1. Top 20 candidate materials in terms of strong magneto-structural coupling. In this table we have included compounds with 20% or less magnetic moment flips.

| Compound | Max. Force difference | Average force |
|---|---|---|
| $Fe_3O_4$ | 1.5156 | 0.5700 |
| $\varepsilon$-$Fe_2O_3$ | 1.4123 | 0.5270 |
| $Fe_4N$ | 0.6483 | 0.5128 |
| $BaFe_4O_7$ | 1.0437 | 0.4673 |
| $CrMn_2O_4$ | 1.2232 | 0.4660 |
| $Fe_4O_5$ | 0.9972 | 0.4645 |
| $Fe_5Na_3O_9$ | 1.0650 | 0.4581 |
| $Fe_7O_9$ | 1.2761 | 0.4307 |
| CrN | 1.4970 | 0.4257 |
| $Mn_3NRh$ | 1.2630 | 0.4127 |
| $AlFeO_3$ | 1.6101 | 0.4094 |
| $Ca_2Fe_7O_{11}$ | 1.1773 | 0.4064 |
| $FeNaO_2$ | 1.1048 | 0.4054 |
| $Fe_2NaO_3$ | 1.0035 | 0.3801 |
| MnN | 1.1663 | 0.3765 |
| $Fe_2Se_2Cs$ | 1.1018 | 0.3729 |
| $Fe_2Ge_2Pr$ | 0.8272 | 0.3642 |
| $AgFeO_2$ | 0.9335 | 0.3608 |
| $BCe_2Fe_{14}$ | 0.9305 | 0.3566 |
| $Fe_2N$ | 0.9744 | 0.3507 |

*3.4 Top candidates review*

Ferrites are of interest for their magnetoelectric and ferroelectric behavior [52,53]. The top candidate material according to our criteria of large average force acting on the atoms is Magnetite $Fe_3O_4$, the first material known as magnetic to mankind. The ground state is ferrimagnetic with antiparallel spin orientation on tetrahedral to octahedral sites, and has a Curie temperature of 860K. Magnetite undergoes a first-order metal-to-insulator Verwey transition at 120K [54]. Proof of ferroelectric order in magnetite films has been presented [55], making magnetite a multiferroic material. $Fe_3O_4$ has been employed for magnetoelectric composites, incorporated as nanoparticles in a ferroelectric P(VDF-TrFE) polymer host [56]. $Fe_3O_4$ has also been considered for actuators, sensors, and vibration energy harvesting systems for its magnetostrictive properties [57]. Experimentally $Fe_3O_4$ has displayed a magnetostriction of 23 ppm [58], almost the same magnitude as bcc Fe (20 ppm) but with opposite sign. ε-$Fe_2O_3$ occupied the second position in the candidate list, it possesses an orthorhombic crystal structure, and undergoes a magnetic transition from a collinear ferrimagnetic ordering to an incommensurate magnetic state occurring at 100 K, where there is ~30% change of permitivity, suggesting the existence of magnetoelectric coupling. Gich *et al*. have shown that in ε-$Fe_2O_3$ nanoparticles the dielectric permittivity can be modified by application of a magnetic field [59]. ε-Fe2O3 is an attractive material since it presents advantages in terms of control of stoichiometry and stable chemical composition. $BaFe_4O_7$ occupies the fourth position. Two different polymorphs have been reported, hexagonal antiferromagnetic $BaFe_4O_7$ with a transition temperature of 945K [60], and trigonal canted antiferromagnetic $BaFe_4O_7$ with a transition temperature of 850K [52].

The corrosion and wear resistance of iron nitrides make them important for the behavior of steels, relevant for coatings and mechanical applications, but also their magnetic properties are of technological interest. FCC structured $Fe_4N$ has been considered for high-density recording materials, spintronics, and as a low-cost core material for transformers [61,62]. The average magnetic moment per iron atom in $Fe_4N$ is close to the value of bcc alpha iron of $2.22\mu_B$ [63]. The magnetic structure of $Fe_4N$ is ferromagnetic with a Curie point at 761K [64].

The crystal structure of $Fe_4N$ has been under debate, the structure reported in the COD and thus used here places the N atom tetrahedrally coordinated while the structure reported in other databases places N in a octahedron environment. DFT calculations indicate that the octahedron environment is the most stable structure. This might have an effect on the forces and change the position of Fe4N on our result list. This illustrates the dependance on the database that is an inherent characteristic of any high throughput study.

CrN has been widely used for coatings in cutting tool technologies for its hardness, but also displays strong magneto-structural coupling. As mentioned in the introduction CrN undergoes a magnetic and structural transition at the critical temperature from a cubic paramagnetic to an orthorhombic antiferromagnetic phase. CrN displays an anomalous thermal conductivity [65], with an almost constant value above the magnetic critical temperature due to the coupling between spin and lattice degrees of freedom [18] , and mechanical properties such as the bulk modulus depend strongly on magnetic effects [1,2].

Transition metal chalcogenides have been of research interest for their potential applications in thermoelectrics [66-68], nonlinear optics [69], and superconductivity [70,71]. Chromites have been revealed to have spontaneous electrical polarization at temperatures lower than ferrimagnetic transitions [51]. This effect makes $Mn_2CrO_4$ and the more famous $CoCr_2O_4$ attractive in spintronic applications. The transition-metal spinels of $AB_2X_4$ general formula form a large class of materials with very rich orbital/charge/spin ordering phenomena [72]. Recently, magnetic spinel oxides have been reported to exhibit dielectric anomaly and ferroelectric properties, both related to their magnetism [73-76].

Manganese based antiperovskites display several attractive properties such as magneto volume effects [77-79], magnetostriction [80,81], magnetocaloric effects [82,83], and magnetoresistance [84,85]. The spin-lattice coupling displayed by $Mn_3RhN$ and Mn-based antiperovskites nitrides are of particular interest in emerging non-volatile magnetic random-access memory (MRAM) devices, through their piezomagnetism by probing and controlling the AFM ordering via the strain-induced magnetic moment [4].

$Na_3Fe_5O_9$ has monoclinic structure and is antiferromagnetic with a Néel temperature of 381 K [86]. $NaFe_2O_3$ is antiferromagnetic with a Néel temperature of 230K, and displays charge ordering at 250K. $NaFe_2O_3$ has been intensively studied for its potential to exhibit exotic ground states realized by the influence of spin frustration [87]. $NaFeO_2$ crystallizes in different phases called $\alpha$, $\beta$, and $\gamma$. The $\alpha$ and $\beta$ phases have been studied for their ferroelectric properties [88]. $\beta$-$NaFeO_2$ is a ferroelectric and weak ferromagnetic compound with ordering temperatures above room temperature. This compound has the potential to become a top candidate for applications in spintronics [88], and memory elements where electrical writing and magnetic reading would present a good balance with respect to energy consumption. $\alpha$-$NaFeO_2$ has a rock-salt crystal structure with space group R3m, with two magnetic phase transitions at 10.5 and 5 K in zero magnetic field.

## 4. Conclusions

In this work, we have proposed an efficient screening procedure to identify potential materials with strong spin-lattice coupling using a descriptor based on the force acting on each atom caused by different local magnetic environments in a disordered magnetic supercell. The method focuses on the underlying atomistic and electronic structure origin in common for several magneto-structural coupling phenomena. The proposed method is not computationally demanding, allowing to probe a large number of compounds. We show that it successfully identifies materials reported for their magneto-structural coupling related phenomena. We provide a list of candidates displaying very strong coupling, among which we find compounds that are already known to display magnetostriction, magnetovolume, magnetocaloric, and magnetoelectric phenomena, while others have not yet been consider for such applications. The top candidate according to our descriptor is $Fe_3O_4$. Among the top candidates we found lesser investigated compounds such as $CrMn_2O_4$ and $Ca_2Fe_7O_{11}$. This work is also a starting point in establishing a materials scale for ranking the strength of their atomistic spin-lattice coupling, independent of any particular macroscopic manifestation. As a continuation of this work, we suggest that the highest ranked candidates are considered for applications where the coupling of spin and lattice degrees of freedom is the key or has been suggested to be beneficial and subjected for in-depth theoretical and experimental investigations.

## Acknowledgements

B.A. acknowledges financial support from the Swedish Government Strategic Research Area in Materials Science on Functional Materials at Linköping University, Faculty Grant SFOMatLiU No. 2009 00971, as well as support from the Swedish Foundation for Strategic Research through the Future Research Leaders 6 program, FFL 15-0290, from the Swedish Research Council (VR) through the grant 2019-05403, and from the Knut and Alice Wallenberg Foundation (Wallenberg Scholar Grant No. KAW-2018.0194). R.A. acknowledges financial support from the Swedish e-Science Research Centre (SeRC) and the Swedish Research Council (VR) project no. 2016-04810 and 2020-05402. The computations were enabled by resources provided by the Swedish National Infrastructure for Computing (SNIC) at NSC partially funded by the Swedish Research Council through grant agreement no. 2018-05973.

## References

[1] B. Alling, T. Marten, and I. A. Abrikosov, *Questionable collapse of the bulk modulus in CrN*, Nat. Mater., **9**, 283 (2010).

[2] F. Rivadulla, M. Bañobre-López, C. X. Quintela, A. Piñeiro, V. Pardo, D. Baldomir, M. A. López-Quintela, J. Rivas, C. A. Ramos, and H. Salva, *Reduction of the bulk modulus at high pressure in CrN*, Nat. Mater., **8**, 947 (2009).

[3] B. L. Gyorffy, A. J. Pindor, J. Staunton, G. M. Stocks, and H. Winter, *A first-principles theory of ferromagnetic phase transitions in metals*, J. Phys. F Met. Phys., **15**, 1337 (1985).

[4] J. Zemen, Z. Gercsi, and K. Sandeman, *Piezomagnetism as a counterpart of the magnetovolume effect in magnetically frustrated Mn-based antiperovskite nitrides*, Phys. Rev. B, **96**, 024451 (2017).

[5] E. R. Callen and H. B. J. P. R. Callen, *Static magnetoelastic coupling in cubic crystals*, Phys. Rev., **129**, 578 (1963).

[6] E. Hristoforou and A. Ktena, *Magnetostriction and magnetostrictive materials for sensing applications*, J. Magn. Magn. Mater., **316**, 372 (2007).

[7] Y. Kim and Y. Y. Kim, *A novel Terfenol-D transducer for guided-wave inspection of a rotating shaft*, Sens. Actuator A Phys., **133**, 447 (2007).

[8] M. Fiebig, *Revival of the magnetoelectric effect*, J. Phys. D Appl. Phys., **38**, R123 (2005).

[9] J. Ma, J. Hu, Z. Li, and C. W. Nan, *Recent progress in multiferroic magnetoelectric composites: from bulk to thin films*, Adv. Mater., **23**, 1062 (2011).

[10] A. P. Pyatakov and A. K. Zvezdin, *Magnetoelectric and multiferroic media*, Physics-Uspekhi, **55**, 557 (2012).

[11] J. Ryu, S. Priya, K. Uchino, and H.-E. Kim, *Magnetoelectric effect in composites of magnetostrictive and piezoelectric materials*, J. Electroceramics, **8**, 107 (2002).

[12] N. A. Spaldin and R. Ramesh, *Advances in magnetoelectric multiferroics*, Nat. Mater., **18**, 203 (2019).

[13] T. Varga, A. Kumar, E. Vlahos, S. Denev, M. Park, S. Hong, T. Sanehira, Y. Wang, C. Fennie, and S. Streiffer, *Coexistence of weak ferromagnetism and ferroelectricity in the high pressure $LiNbO_3$-type phase of $FeTiO_3$*, Phys. Rev. Lett., **103**, 047601 (2009).

[14] J. Liu, T. Gottschall, K. P. Skokov, J. D. Moore, and O. Gutfleisch, *Giant magnetocaloric effect driven by structural transitions*, Nat. Mater., **11**, 620 (2012).

[15] A. Tishin, Y. Spichkin, V. Zverev, and P. Egolf, *A review and new perspectives for the magnetocaloric effect: New materials and local heating and cooling inside the human body*, Int. J. Refrig., **68**, 177 (2016).

[16] A. Kitanovski and P. W. Egolf, *Innovative ideas for future research on magnetocaloric technologies*, Int. J. Refrig., **33**, 449 (2010).

[17] A. M. Tishin, Y. I. Spichkin, V. I. Zverev, and P. W. Egolf, *A review and new perspectives for the magnetocaloric effect: New materials and local heating and cooling inside the human body*, Int. J. Refrig., **68**, 177 (2016).

[18] I. Stockem, A. Bergman, A. Glensk, T. Hickel, F. Körmann, B. Grabowski, J. Neugebauer, and B. Alling, *Anomalous Phonon Lifetime Shortening in Paramagnetic CrN Caused by Spin-Lattice Coupling: A Combined Spin and Ab Initio Molecular Dynamics Study*, Phys. Rev. Lett., **121**, 125902 (2018).

[19] Y. Grosu, A. Faik, I. Ortega-Fernández, and B. D'Aguanno, *Natural Magnetite for thermal energy storage: Excellent thermophysical properties, reversible latent heat transition and controlled thermal conductivity*, Sol. Energy Mater Sol. Cells, **161**, 170 (2017).

[20] L. Casillas-Trujillo, D. Gambino, L. Ventelon, and B. Alling, *Screw dislocation core structure in the paramagnetic state of bcc iron from first-principles calculations*, Phys. Rev. B, **102**, 094420 (2020).
[21] G. Ceder and K. Persson, *How supercomputers will yield a golden age of materials science*, Sci. Am., **309**, 1 (2013).
[22] S. Curtarolo, G. L. W. Hart, M. B. Nardelli, N. Mingo, S. Sanvito, and O. Levy, *The high-throughput highway to computational materials design*, Nat. Mater., **12**, 191 (2013).
[23] G. L. Hart, S. Curtarolo, T. B. Massalski, and O. Levy, *Comprehensive search for new phases and compounds in binary alloy systems based on platinum-group metals, using a computational first-principles approach*, Phys. Rev. X, **3**, 041035 (2013).
[24] K. Yang, W. Setyawan, S. Wang, M. B. Nardelli, and S. Curtarolo, *A search model for topological insulators with high-throughput robustness descriptors*, Nat. Mater., **11**, 614 (2012).
[25] D. Goll, R. Loeffler, J. Herbst, R. Karimi, and G. Schneider, *High-throughput search for new permanent magnet materials*, J. Condens. Matter Phys., **26**, 064208 (2014).
[26] M. Jaime, A. Saúl, M. Salamon, V. Zapf, N. Harrison, T. Durakiewicz, J. Lashley, D. Andersson, C. Stanek, and J. Smith, *Piezomagnetism and magnetoelastic memory in uranium dioxide*, Nat. Commun., **8**, 99 (2017).
[27] M. L. Green, I. Takeuchi, and J. R. Hattrick-Simpers, *Applications of high throughput (combinatorial) methodologies to electronic, magnetic, optical, and energy-related materials*, J. Appl. Phys., **113**, 9_1 (2013).
[28] A. G. Kusne, T. Gao, A. Mehta, L. Ke, M. C. Nguyen, K.-M. Ho, V. Antropov, C.-Z. Wang, M. J. Kramer, and C. Long, *On-the-fly machine-learning for high-throughput experiments: search for rare-earth-free permanent magnets*, Sci. Rep., **4**, 6367 (2014).
[29] M. K. Horton, J. H. Montoya, M. Liu, and K. A. Persson, *High-throughput prediction of the ground-state collinear magnetic order of inorganic materials using Density Functional Theory*, Npj Comput. Mater., **5**, 2 (2019).
[30] J. D. Bocarsly, E. E. Levin, C. A. Garcia, K. Schwennicke, S. D. Wilson, and R. Seshadri, *A simple computational proxy for screening magnetocaloric compounds*, Chem. Mater., **29**, 1613 (2017).
[31] The High-Throughput Toolkit (httk), R. Armiento et al., http://httk.openmaterialsdb.se/.
[32] F. Körmann, A. Dick, B. Grabowski, T. Hickel, and J. Neugebauer, *Atomic forces at finite magnetic temperatures: Phonons in paramagnetic iron*, Phys. Rev. B, **85**, 125104 (2012).
[33] B. Alling, *Theory of the ferromagnetism in* $Ti_{1-x}Cr_xN$ *solid solutions*, Phys. Rev. B, **82**, 054408 (2010).
[34] B. Alling, T. Marten, and I. Abrikosov, *Effect of magnetic disorder and strong electron correlations on the thermodynamics of CrN*, Phys. Rev. B, **82**, 184430 (2010).
[35] A. Zunger, S.-H. Wei, L. Ferreira, and J. E. Bernard, *Special quasirandom structures*, Phys. Rev. Lett., **65**, 353 (1990).
[36] I. A. Abrikosov, A. V. Ponomareva, P. Steneteg, S. A. Barannikova, and B. Alling, *Recent progress in simulations of the paramagnetic state of magnetic materials*, Curr. Opin. Solid State Mater. Sci.
, **20**, 85 (2016).
[37] D. Gambino and B. Alling, *Lattice relaxations in disordered Fe-based materials in the paramagnetic state from first principles*, Phys. Rev. B, **98**, 064105 (2018).

[38] A. Filippetti and N. A. Hill, *Magnetic stress as a driving force of structural distortions: the case of CrN*, Phys. Rev. Lett., **85**, 5166 (2000).
[39] J. D. Bocarsly, E. E. Levin, S. A. Humphrey, T. Faske, W. Donner, S. D. Wilson, and R. Seshadri, *Magnetostructural Coupling Drives Magnetocaloric Behavior: The Case of MnB versus FeB*, Chem. Mater., **31**, 4873 (2019).
[40] J. Browne, P. Liddell, R. Street, and T. Mills, *An investigation of the antiferromagnetic transition of CrN*, Phys. Status Solidi A, **1**, 715 (1970).
[41] L. Corliss, N. Elliott, and J. Hastings, *Antiferromagnetic structure of CrN*, Physical Review, **117**, 929 (1960).
[42] P. S. Herle, M. Hegde, N. Vasathacharya, S. Philip, M. R. Rao, and T. Sripathi, *Synthesis of TiN, VN, and CrN from ammonolysis of $TiS_2$, $VS_2$, and $Cr_2S_3$*, J Solid State Chem., **134**, 120 (1997).
[43] R. Ibberson and R. Cywinski, *The magnetic and structural transitions in CrN and (CrMo) N*, Physica B Condens. Matter, **180**, 329 (1992).
[44] S. Gražulis, D. Chateigner, R. T. Downs, A. Yokochi, M. Quirós, L. Lutterotti, E. Manakova, J. Butkus, P. Moeck, and A. Le Bail, *Crystallography Open Database–an open-access collection of crystal structures*, J. Appl. Crystallogr., **42**, 726 (2009).
[45] G. Kresse and J. Furthmüller, *Efficiency of ab-initio total energy calculations for metals and semiconductors using a plane-wave basis set*, Comput. Mater. Sci., **6**, 15 (1996).
[46] G. Kresse and J. Furthmüller, *Efficient iterative schemes for ab initio total-energy calculations using a plane-wave basis set*, Phys. Rev. B, **54**, 11169 (1996).
[47] J. P. Perdew, K. Burke, and M. Ernzerhof, *Generalized Gradient Approximation Made Simple*, Phys. Rev. Lett., **77**, 3865 (1996).
[48] P. E. Blöchl, *Projector augmented-wave method*, Phys. Rev. B, **50**, 17953 (1994).
[49] See Supplemental Material at [URL] for full results list.
[50] L. Y. Gavrilova, T. Aksenova, and V. Cherepanov, *Phase equilibria and crystal structures of complex oxides in systems La-M-Fe-O (M= Ca or Sr)*, Russ. J. Inorg. Chem., **53**, 953 (2008).
[51] Y.-C. Jhuang, K. Kuo, and G. Chern, *Structural and magnetic characterizations of $Mn_2CrO_4$ and $MnCr_2O_4$ films on MgO (001) and $SrTiO_3$ (001) substrates by molecular beam epitaxy*, J. Appl. Phys., **109**, 07D714 (2011).
[52] T. Ferreira, G. Morrison, W. M. Chance, S. Calder, M. D. Smith, and H.-C. zur Loye, *$BaFe_4O_7$ and $K_{0.22}Ba_{0.89}Fe_4O_7$: Canted Antiferromagnetic Diferrites with Exceptionally High Magnetic Ordering Temperatures*, Chem. Mater., **29**, 2689 (2017).
[53] G. Zhang, J. Hou, M. Zhu, G. Huang, D. Li, Y. Fang, and T. Zeng, *Visible-light photovoltaic effect in high-temperature ferroelectric $BaFe_4O_7$*, J. Mater. Chem. C, **8**, 16234 (2020).
[54] F. Walz, *The Verwey transition-a topical review*, J. Phys. Condens. Matter., **14**, R285 (2002).
[55] M. Alexe, M. Ziese, D. Hesse, P. Esquinazi, K. Yamauchi, T. Fukushima, S. Picozzi, and U. Gösele, *Ferroelectric switching in multiferroic magnetite ($Fe_3O_4$) thin films*, Adv. Mater., **21**, 4452 (2009).
[56] R. Belouadah, L. Seveyrat, D. Guyomar, B. Guiffard, and F. Belhora, *Magnetoelectric coupling in $Fe_3O_4$/P (VDF-TrFE) nanocomposites*, Sens. Actuator A Phys., **247**, 298 (2016).
[57] D. Odkhuu, P. Taivansaikhan, W. S. Yun, and S. C. Hong, *A first-principles study of magnetostrictions of $Fe_3O_4$ and $CoFe_2O_4$*, J. Appl. Phys., **115**, 17A916 (2014).
[58] R. M. Bozorth, E. F. Tilden, and A. J. Williams, *Anisotropy and Magnetostriction of Some Ferrites*, Phys. Rev., **99**, 1788 (1955).

[59] M. Gich, C. Frontera, A. Roig, J. Fontcuberta, E. Molins, N. Bellido, C. Simon, and C. Fleta, *Magnetoelectric coupling in ε-$Fe_2O_3$ nanoparticles*, Nanotechnology, **17**, 687 (2006).
[60] S. Okamoto, H. Sekizawa, and S. I. Okamoto, *Hydrothermal synthesis, structure and magnetic properties of barium diferrite*, J. Phys. Chem. Solids, **36**, 591 (1975).
[61] T. Monson, B. Zheng, Y. Zhou, E. J. Lavernia, C. J. Pearce, and S. Atcitty, gamma'-Fe4N a new soft magnetic material for inductors and motors, Report No. (No. SAND2016-8042C). , 2016.Sandia National Lab.(SNL-NM), Albuquerque, NM (United States); Sandia National Laboratories, Livermore, CA, 2016.
[62] T. E. Stevens, C. J. Pearce, M. A. Rodriguez, S. Dickens, B. B. McKenzie, S. Atcitty, and T. Monson, Synthesis of gamma'-$Fe_4N$ a new soft magnetic material for inductors and transformers, Report No. No. SAND2018-5099C, 2018.Sandia National Lab.(SNL-NM), Albuquerque, NM (United States)
[63] S. Chen, S. Jin, T. Tiefel, Y. Hsieh, E. Gyorgy, and D. Johnson Jr, *Magnetic properties and microstructure of $Fe_4N$ and (Fe, Ni)$_4$N*, J. Appl. Phys., **70**, 6247 (1991).
[64] B. Frazer, *Magnetic structure of $Fe_4N$*, Phys. Rev., **112**, 751 (1958).
[65] O. Jankovský, D. Sedmidubský, Š. Huber, P. Šimek, and Z. Sofer, *Synthesis, magnetic and transport properties of oxygen-free CrN ceramics*, J. Eur. Ceram. Soc., **34**, 4131 (2014).
[66] M. G. Kanatzidis, *Nanostructured thermoelectrics: the new paradigm?*, Chem. Mater., **22**, 648 (2009).
[67] A. Maignan, E. Guilmeau, F. Gascoin, Y. Bréard, and V. Hardy, *Revisiting some chalcogenides for thermoelectricity*, Sci. Technol. Adv. Mater., **13**, 053003 (2012).
[68] G. Tan, L.-D. Zhao, and M. G. Kanatzidis, *Rationally designing high-performance bulk thermoelectric materials*, Chem. Rev., **116**, 12123 (2016).
[69] I. Chung and M. G. Kanatzidis, *Metal chalcogenides: a rich source of nonlinear optical materials*, Chem. Mater., **26**, 849 (2013).
[70] D. C. Johnston, *The puzzle of high temperature superconductivity in layered iron pnictides and chalcogenides*, Adv. Phys., **59**, 803 (2010).
[71] H. K. Vivanco and E. E. Rodriguez, *The intercalation chemistry of layered iron chalcogenide superconductors*, J Solid State Chem., **242**, 3 (2016).
[72] K. Singh, A. Maignan, C. Simon, and C. Martin, *$FeCr_2O_4$ and $CoCr_2O_4$ spinels: Multiferroicity in the collinear magnetic state?*, Appl. Phys. Lett., **99**, 172903 (2011).
[73] Y. J. Choi, J. Okamoto, D. Huang, K. Chao, H. Lin, C. Chen, M. Van Veenendaal, T. Kaplan, and S.-W. Cheong, *Thermally or magnetically induced polarization reversal in the multiferroic $CoCr_2O_4$*, Phys. Rev. Lett., **102**, 067601 (2009).
[74] G. Lawes, B. Melot, K. Page, C. Ederer, M. Hayward, T. Proffen, and R. Seshadri, *Dielectric anomalies and spiral magnetic order in $CoCr_2O_4$*, Phys. Rev. B, **74**, 024413 (2006).
[75] N. Mufti, A. Nugroho, G. Blake, and T. Palstra, *Magnetodielectric coupling in frustrated spin systems: the spinels $MCr_2O_4$ (M= Mn, Co and Ni)*, J. Phys. Condens. Matter., **22**, 075902 (2010).
[76] Y. Yamasaki, S. Miyasaka, Y. Kaneko, J.-P. He, T. Arima, and Y. Tokura, *Magnetic reversal of the ferroelectric polarization in a multiferroic spinel oxide*, Phys. Rev. Lett., **96**, 207204 (2006).
[77] D. Fruchart and E. F. Bertaut, *Magnetic studies of the metallic perovskite-type compounds of manganese*, J. Phys. Soc. Jpn., **44**, 781 (1978).
[78] T. Kaneko, T. Kanomata, and K. Shirakawa, *Pressure effect on the magnetic transition temperatures in the intermetallic compounds* $Mn_3MC$ *(M= Ga, Zn and Sn)*, J. Phys. Soc. Jpn., **56**, 4047 (1987).

[79] K. Takenaka, M. Ichigo, T. Hamada, A. Ozawa, T. Shibayama, T. Inagaki, and K. Asano, *Magnetovolume effects in manganese nitrides with antiperovskite structure*, Sci. Technol. Adv. Mater., **15**, 015009 (2014).
[80] K. Asano, K. Koyama, and K. Takenaka, *Magnetostriction in $Mn_3CuN$*, Appl. Phys. Lett., **92**, 161909 (2008).
[81] K. Takenaka, T. Shibayama, D. Kasugai, and T. Shimizu, *Giant Field-Induced Distortion in $Mn_3SbN$ at Room Temperature*, Jpn. J. Appl. Phys., **51**, 043001 (2012).
[82] T. Tohei, H. Wada, and T. Kanomata, *Negative magnetocaloric effect at the antiferromagnetic to ferromagnetic transition of $Mn_3GaC$*, J. Appl. Phys., **94**, 1800 (2003).
[83] B. Wang, P. Tong, Y. Sun, X. Luo, X. Zhu, G. Li, X. Zhu, S. Zhang, Z. Yang, and W. Song, *Large magnetic entropy change near room temperature in antiperovskite $SnCMn_3$*, EPL (Europhysics Letters), **85**, 47004 (2009).
[84] K. Kamishima, T. Goto, H. Nakagawa, N. Miura, M. Ohashi, N. Mori, T. Sasaki, and T. Kanomata, *Giant magnetoresistance in the intermetallic compound $Mn_3GaC$*, Phys. Rev. B, **63**, 024426 (2000).
[85] B. S. Wang, P. Tong, Y. P. Sun, L. J. Li, W. Tang, W. J. Lu, X. B. Zhu, Z. R. Yang, and W. H. Song, *Enhanced giant magnetoresistance in Ni-doped antipervoskite compounds $GaCMn_{3-x}Ni_x(x=0.05,0.10)$*, Appl. Phys. Lett., **95**, 222509 (2009).
[86] C. Romers, C. Rooymans, and R. De Graaf, *The preparation, crystal structure and magnetic properties of $Na_3Fe_5O_9$*, Acta Crystallogr., **22**, 766 (1967).
[87] S. Kobayashi, H. Ueda, C. Michioka, K. Yoshimura, S. Nakamura, T. Katsufuji, and H. Sawa, *Anomalous double-stripe charge ordering in $\beta$−$NaFe_2O_3$ with double triangular layers consisting of almost perfect regular Fe 4 tetrahedra*, Phys. Rev. Mater., **2**, 054402 (2018).
[88] M. Viret, D. Rubi, D. Colson, D. Lebeugle, A. Forget, P. Bonville, G. Dhalenne, R. Saint-Martin, G. Andre, and F. Ott, *$\beta$-$NaFeO_2$, a new room-temperature multiferroic material*, Mater. Res. Bull, **47**, 2294 (2012).

Appendix A: Candidate list

Max Δforce: maximum force difference between different DLM configurations.
Avg. Δforce: average of the force difference between different DLM configurations
Avg. force: Average force acting on each atom.
Spin flip %: percentage of magnetic atoms that have changed sign from their initial setting.

| Compound | Max Δforce | Avg. Δforce | Avg. force | Spin flip % |
|---|---|---|---|---|
| $Fe_3O_4$ | 1.5156 | 0.6936 | 0.5700 | 8% |
| $Fe_2O_3$ | 1.4123 | 0.6477 | 0.5270 | 19% |
| $Fe_4N$ | 0.6483 | 0.4064 | 0.5128 | 0% |
| $BaFe_4O_7$ | 1.0437 | 0.4012 | 0.4673 | 19% |
| $CrMn_2O_4$ | 1.2232 | 0.4193 | 0.4660 | 0% |
| $Fe_4O_5$ | 0.9972 | 0.4331 | 0.4645 | 12% |
| $Fe_5Na_3O_9$ | 1.0650 | 0.4869 | 0.4581 | 10% |
| $Fe_2S_4Si$ | 1.8489 | 0.4612 | 0.4491 | >20% |
| FeN | 0.8733 | 0.5797 | 0.4440 | >20% |
| $CuFeO_2$ | 1.3778 | 0.5917 | 0.4312 | >20% |
| $Fe_7O_9$ | 1.2761 | 0.4202 | 0.4307 | 14% |
| CrN | 1.4970 | 0.6460 | 0.4257 | 3% |
| $Mn_3NRh$ | 1.2630 | 0.5677 | 0.4127 | 12% |
| $AlFeO_3$ | 1.6101 | 0.4187 | 0.4094 | 12% |
| $Ca_2Fe_7O_{11}$ | 1.1773 | 0.4259 | 0.4064 | 7% |
| $FeNaO_2$ | 1.1048 | 0.4123 | 0.4054 | 12% |
| $Er_2Fe_{17}$ | 1.0150 | 0.4182 | 0.3897 | >20% |
| $Fe_2NaO_3$ | 1.0035 | 0.3961 | 0.3801 | 12% |
| $FeTiO_3$ | 1.3574 | 0.5314 | 0.3784 | >20% |
| MnN | 1.1663 | 0.5621 | 0.3765 | 0% |
| $FeTmO_3$ | 1.3242 | 0.4776 | 0.3765 | >20% |
| $CsFe_2Se_2$ | 1.1018 | 0.3221 | 0.3729 | 3% |
| $BFe_{14}Tb_2$ | 0.9145 | 0.3452 | 0.3714 | >20% |
| $Fe_2Ge_2Pr$ | 0.8272 | 0.2356 | 0.3642 | 0% |
| $AgFeO_2$ | 0.9335 | 0.3337 | 0.3608 | 17% |
| $BCe_2Fe_{14}$ | 0.9305 | 0.3452 | 0.3566 | 5% |
| $BFe_{14}Ho_2$ | 0.7758 | 0.3145 | 0.3522 | >20% |
| $Fe_2N$ | 0.9744 | 0.3993 | 0.3507 | 0% |
| $BFe_{14}Gd_2$ | 0.8913 | 0.3743 | 0.3498 | >20% |
| $Fe_{17}Nd_2$ | 0.7767 | 0.3241 | 0.3495 | 12% |
| $BFe_{14}Pr_2$ | 0.6121 | 0.2764 | 0.3467 | 5% |
| $Mn_3NNi$ | 1.1572 | 0.3895 | 0.3456 | 12% |
| $FeO_2$ | 0.6617 | 0.3657 | 0.3434 | 6% |
| $Fe_{17}Pr_2$ | 0.9366 | 0.3364 | 0.3402 | 8% |
| $BFe_{14}Nd_2$ | 0.9413 | 0.3486 | 0.3386 | 5% |
| $As_2EuFe_2$ | 0.8354 | 0.3146 | 0.3370 | >20% |
| $Fe_3Ge_2O_8$ | 1.9544 | 0.5219 | 0.3350 | 0% |
| $F_6Fe_2Li$ | 0.7313 | 0.2961 | 0.3320 | 6% |
| $BiFeO_3$ | 1.0374 | 0.3306 | 0.3282 | 6% |
| $Fe_3Sn$ | 0.8956 | 0.4349 | 0.3276 | 0% |
| $Co_5Pr$ | 1.2308 | 0.5569 | 0.3271 | 10% |
| $Ce_2Fe_{17}$ | 1.1947 | 0.3396 | 0.3270 | 14% |
| $As_3CaFe_5$ | 0.7692 | 0.3204 | 0.3266 | 10% |
| $Cr_2S_3$ | 1.4882 | 0.3358 | 0.3237 | 12% |
| $BaFe_{12}O_{19}$ | 1.1468 | 0.4286 | 0.3232 | 8% |
| $Mn_2P$ | 0.6437 | 0.2416 | 0.3230 | >20% |
| $BaFe_2Se_3$ | 0.4277 | 0.2085 | 0.3225 | 0% |
| AsLiMn | 0.6638 | 0.2389 | 0.3223 | 0% |
| $Mn_3NPt$ | 0.8553 | 0.3399 | 0.3186 | >20% |
| $Fe_2KSe_2$ | 1.2490 | 0.4189 | 0.3185 | 9% |
| GaMn | 0.6029 | 0.2455 | 0.3162 | 3% |
| MnO | 1.0407 | 0.4517 | 0.3151 | 6% |
| FeS | 0.8795 | 0.3006 | 0.3128 | >20% |
| $CrO_2$ | 0.6884 | 0.3354 | 0.3124 | 12% |
| GdGeMn | 0.7957 | 0.2854 | 0.3113 | >20% |
| $Fe_3Ga$ | 0.8129 | 0.3941 | 0.3104 | 0% |
| $Fe_6Ge_6Mg$ | 1.5237 | 0.4250 | 0.3093 | 0% |
| $Fe_2MgO_4$ | 1.1102 | 0.4014 | 0.3083 | >20% |
| $Mn_2TaO_3$ | 1.1383 | 0.4478 | 0.3066 | 12% |
| $FeLuO_3$ | 0.8311 | 0.4125 | 0.3053 | >20% |
| $C_2Fe_5$ | 0.5395 | 0.3156 | 0.3051 | 20% |
| $C_2Fe_{11}Th$ | 0.4561 | 0.2422 | 0.3031 | 18% |
| $Fe_3N$ | 0.7638 | 0.3626 | 0.3015 | 8% |
| $Fe_{13}PbPr_6$ | 0.6271 | 0.2996 | 0.2998 | 15% |
| $FeLaO_3$ | 1.2549 | 0.3574 | 0.2987 | 12% |
| $Fe_4Ge_2Yb$ | 0.5811 | 0.2753 | 0.2986 | >20% |
| $BMn_3O_5$ | 1.6164 | 0.5030 | 0.2986 | 17% |
| ClFeO | 0.8810 | 0.2402 | 0.2904 | 6% |
| $MnYbO_3$ | 1.2920 | 0.4480 | 0.2871 | >20% |
| $Mn_3NPd$ | 1.2253 | 0.4682 | 0.2859 | 17% |
| $AsMn_2$ | 0.4816 | 0.2774 | 0.2858 | 0% |
| $C_3Fe_{17}Tb$ | 0.7178 | 0.3661 | 0.2852 | 18% |
| $CoF_3$ | 1.8717 | 0.4938 | 0.2841 | >20% |
| FeSn | 1.0993 | 0.4117 | 0.2838 | 0% |
| $Cr_2O_3$ | 0.7395 | 0.2072 | 0.2834 | 0% |
| $BFe_3O_5$ | 1.0923 | 0.3641 | 0.2817 | 8% |
| $Fe_{17}Gd_2$ | 0.5462 | 0.2746 | 0.2808 | 12% |
| FeSe | 1.5371 | 0.4515 | 0.2806 | 6% |
| $CFe_2$ | 0.8709 | 0.4057 | 0.2781 | 6% |
| $Cr_3Se_4$ | 0.5662 | 0.2589 | 0.2781 | 4% |
| $Fe_6Ge_6Hf$ | 0.7029 | 0.2925 | 0.2764 | >20% |
| GeMnY | 1.0570 | 0.4775 | 0.2758 | >20% |
| $Fe_2S_4Tl_3$ | 0.7438 | 0.2470 | 0.2756 | 19% |
| $BMn_2O_4$ | 1.8954 | 0.4005 | 0.2739 | 6% |
| $Fe_3Pd$ | 0.8122 | 0.4127 | 0.2722 | 12% |
| $Mn_3Pt$ | 1.4781 | 0.4647 | 0.2715 | >20% |
| $FeYO_3$ | 0.9520 | 0.2942 | 0.2708 | >20% |
| $Co_5Nd$ | 0.9038 | 0.3748 | 0.2680 | 5% |
| $Fe_3O_7P$ | 1.3283 | 0.3537 | 0.2669 | 6% |
| $Ca_2Fe_2O_5$ | 0.7986 | 0.3324 | 0.2662 | >20% |
| $Mn_2O_3$ | 0.8149 | 0.3486 | 0.2657 | 6% |
| CoFe | 0.7987 | 0.4876 | 0.2650 | >20% |
| $DyMn_2O_5$ | 1.3478 | 0.3426 | 0.2638 | >20% |
| $Mn_2Si_2U$ | 0.6460 | 0.2581 | 0.2632 | >20% |
| $ErFeO_3$ | 1.0391 | 0.3764 | 0.2618 | >20% |
| $MnO_2$ | 0.5847 | 0.2886 | 0.2615 | 12% |
| $GeMn_3$ | 0.7271 | 0.3862 | 0.2611 | 0% |
| $Fe_2Rb_3Se_4$ | 0.8090 | 0.2659 | 0.2596 | 0% |
| $MnSnO_3$ | 0.8790 | 0.2692 | 0.2589 | 17% |
| $FeKO_2$ | 1.1311 | 0.3569 | 0.2517 | 6% |
| $CGeMn_3$ | 0.8971 | 0.3825 | 0.2515 | 8% |
| $GdMnO_3$ | 1.0412 | 0.2617 | 0.2503 | >20% |
| $Fe_3Th$ | 0.8989 | 0.3003 | 0.2490 | >20% |
| $Fe_2Na_3S_4$ | 0.8480 | 0.3082 | 0.2489 | 19% |
| $As_2Mn_2Yb$ | 0.5169 | 0.1872 | 0.2488 | 0% |
| $C_3Fe_{17}Th_2$ | 0.6231 | 0.2740 | 0.2487 | >20% |
| $BFe_2O_4$ | 0.9322 | 0.3275 | 0.2482 | 6% |
| FeGa | 1.6519 | 0.3270 | 0.2468 | 15% |
| GeMnTm | 1.1275 | 0.2782 | 0.2433 | >20% |
| AsMnNa | 0.8746 | 0.2837 | 0.2433 | 0% |
| $Fe_6Ge_6Sc$ | 0.7297 | 0.4307 | 0.2413 | 17% |
| CrS | 0.8436 | 0.3690 | 0.2413 | 0% |
| $BaMnO_3$ | 2.3042 | 0.2566 | 0.2404 | 12% |
| AsCrRh | 0.4893 | 0.2147 | 0.2378 | >20% |
| CoO | 0.9038 | 0.3488 | 0.2366 | 6% |
| $Fe_3Pt$ | 0.6175 | 0.3019 | 0.2363 | 17% |
| $Fe_2Sn$ | 0.7504 | 0.3364 | 0.2340 | 0% |
| AlGeMn | 0.7997 | 0.2517 | 0.2337 | 6% |
| $LuMn_2O_5$ | 1.6255 | 0.2787 | 0.2336 | >20% |
| $CFe_3$ | 0.4542 | 0.2882 | 0.2320 | 0% |
| BMn | 0.6589 | 0.1951 | 0.2314 | 19% |
| $CuFeSe_2$ | 1.0274 | 0.2793 | 0.2308 | >20% |
| $Mn_2O_4Si$ | 1.0673 | 0.3629 | 0.2300 | 19% |
| GaHoMn | 0.8205 | 0.3419 | 0.2295 | >20% |
| $CuFe_2Ge_2$ | 0.5933 | 0.1696 | 0.2285 | 19% |
| $AuMn_2$ | 0.8426 | 0.3030 | 0.2273 | 9% |
| $BiMn_2O_5$ | 1.5149 | 0.2701 | 0.2273 | 12% |
| $As_2EuMn_2$ | 0.6854 | 0.2476 | 0.2253 | 0% |
| $AsFe_2$ | 0.8313 | 0.3333 | 0.2252 | 19% |
| $Fe_6Ge_4Li$ | 0.6878 | 0.2662 | 0.2243 | 17% |
| $MnSmO_3$ | 0.7462 | 0.2989 | 0.2239 | >20% |
| GeHoMn | 0.6459 | 0.3020 | 0.2233 | >20% |
| $BiMnO_3$ | 0.7404 | 0.2640 | 0.2231 | >20% |
| $Fe_5Y$ | 0.9169 | 0.3048 | 0.2216 | >20% |
| $Fe_6Ge_6Li$ | 0.9550 | 0.4083 | 0.2180 | 6% |
| $EuMn_2Sb_2$ | 0.5653 | 0.2107 | 0.2178 | 6% |
| $Fe_3Te_3Tl$ | 0.6462 | 0.2398 | 0.2174 | 8% |
| $FeNdO_3$ | 0.9014 | 0.3592 | 0.2165 | >20% |
| $MnS_2$ | 1.1518 | 0.3405 | 0.2152 | 12% |
| $FeScO_3$ | 1.2515 | 0.3067 | 0.2152 | 8% |
| $Co_5Gd$ | 0.9127 | 0.3058 | 0.2150 | >20% |
| $As_2Mn_2Zn$ | 0.3650 | 0.1752 | 0.2148 | 0% |
| $Mn_2NpSi_2$ | 0.7465 | 0.2994 | 0.2139 | 4% |
| $Fe_5Th$ | 0.6917 | 0.3213 | 0.2133 | 10% |
| $Mn_2O_5Sr_2$ | 1.7810 | 0.2898 | 0.2128 | >20% |
| $Mn_2O_5Yb$ | 0.8180 | 0.2581 | 0.2114 | >20% |
| $Fe_2Th$ | 0.7816 | 0.3059 | 0.2096 | >20% |
| $Ca_2Co_2O_5$ | 1.0360 | 0.3351 | 0.2092 | 19% |
| $Fe_2O_4Si$ | 0.8988 | 0.2842 | 0.2090 | 12% |
| $LiMn_2O_4$ | 0.9619 | 0.2506 | 0.2089 | >20% |
| GeMnSr | 0.7547 | 0.2498 | 0.2086 | 6% |
| CrSb | 0.8094 | 0.3261 | 0.2077 | 0% |
| $Co_5Er$ | 0.7905 | 0.3497 | 0.2075 | >20% |
| $MnNdO_3$ | 0.7876 | 0.2833 | 0.2065 | 12% |
| $Fe_3Ho$ | 0.7196 | 0.2783 | 0.2057 | >20% |
| $Mn_2NaO_4$ | 1.1505 | 0.2304 | 0.2035 | >20% |
| $Cr_5S_6$ | 1.1409 | 0.1897 | 0.2027 | 0% |
| $LaMnO_3$ | 1.0841 | 0.3120 | 0.2016 | 8% |
| GeMnYb | 0.5929 | 0.2153 | 0.2015 | 6% |
| MnP | 0.5052 | 0.2714 | 0.2011 | 19% |
| AsMnRh | 0.6097 | 0.1878 | 0.1985 | >20% |
| $Mn_5Si_2$ | 0.5331 | 0.2114 | 0.1984 | >20% |
| $Fe_3S_4Tl_2$ | 0.8411 | 0.2752 | 0.1983 | >20% |
| DyGeMn | 0.4338 | 0.2070 | 0.1983 | >20% |
| $As_3Co_5O_{12}$ | 0.8131 | 0.2731 | 0.1982 | 0% |
| $CaMn_7O_{12}$ | 0.8912 | 0.3025 | 0.1982 | 14% |
| $Mn_3Sn$ | 0.5025 | 0.2434 | 0.1978 | 0% |
| $Co_5Tb$ | 0.8046 | 0.3063 | 0.1937 | >20% |
| $MnTbO_3$ | 1.1352 | 0.2553 | 0.1935 | >20% |
| $Cr_2Si_2U$ | 0.6380 | 0.2435 | 0.1925 | >20% |
| $Fe_2MnSi$ | 0.6835 | 0.2687 | 0.1919 | >20% |
| MnSiTb | 0.6182 | 0.2894 | 0.1917 | >20% |
| $Fe_5Gd$ | 0.5407 | 0.2360 | 0.1916 | >20% |
| $Cr_2O_6Re$ | 0.4105 | 0.2260 | 0.1916 | >20% |
| MnS | 0.4605 | 0.2432 | 0.1900 | 0% |
| $Mn_2Sb_2Yb$ | 0.5208 | 0.2388 | 0.1886 | 0% |
| $Mn_3Sb$ | 0.5355 | 0.2277 | 0.1875 | 0% |
| BFe | 0.5297 | 0.2450 | 0.1869 | >20% |
| $EuMnO_3$ | 0.6300 | 0.2342 | 0.1861 | >20% |
| MnSiYb | 0.5681 | 0.1941 | 0.1857 | 0% |
| $BCo_{14}Nd_2$ | 0.5666 | 0.2416 | 0.1854 | >20% |
| $Cl_3Mn_8O_{10}$ | 0.6370 | 0.2455 | 0.1850 | 0% |
| $BErFe_4$ | 0.6533 | 0.2286 | 0.1848 | >20% |
| $Fe_5Ho$ | 0.7714 | 0.2784 | 0.1845 | >20% |
| $CoTiO_3$ | 0.7857 | 0.3213 | 0.1830 | >20% |
| $Co_2FeIn$ | 0.9675 | 0.2739 | 0.1825 | >20% |
| $GdMn_2O_5$ | 0.9317 | 0.1997 | 0.1823 | >20% |
| $Fe_5Si_3$ | 0.6953 | 0.2336 | 0.1821 | 10% |
| $Mn_2Sb$ | 0.4302 | 0.2300 | 0.1818 | 0% |
| $Fe_2Ti$ | 0.2635 | 0.1064 | 0.1807 | >20% |
| $EuMn_2P_2$ | 0.6388 | 0.2522 | 0.1800 | 6% |
| $MnPrO_3$ | 0.7856 | 0.2721 | 0.1779 | 12% |
| BaGeMn | 0.5152 | 0.1647 | 0.1776 | 0% |
| $FeSiO_3$ | 1.1786 | 0.2967 | 0.1761 | 19% |
| CoMnSi | 0.3959 | 0.2064 | 0.1761 | >20% |
| $Mn_6Sn_6Tb$ | 0.5066 | 0.2451 | 0.1747 | >20% |
| $Fe_5Nd$ | 0.4831 | 0.2565 | 0.1735 | 20% |
| $DyFe_5$ | 0.6791 | 0.2385 | 0.1727 | 20% |
| $Ge_2Mn_2U$ | 0.5782 | 0.2201 | 0.1716 | 0% |
| $CeFe_5$ | 0.5870 | 0.2109 | 0.1708 | 15% |
| $Cr_3Se_4$ | 0.5267 | 0.2035 | 0.1705 | 4% |
| $Co_5Y$ | 0.5479 | 0.2666 | 0.1703 | >20% |
| FeSb | 0.4235 | 0.1892 | 0.1691 | 12% |
| $F_6Fe_2Rb$ | 0.7016 | 0.2277 | 0.1686 | 19% |
| $GeMn_2O_4$ | 1.0371 | 0.2268 | 0.1685 | 12% |
| $BCo_4Y$ | 0.2514 | 0.1413 | 0.1681 | >20% |
| $Mn_2Si_2Y$ | 0.8423 | 0.2396 | 0.1679 | >20% |
| $CoF_3K$ | 0.8017 | 0.2101 | 0.1676 | 12% |
| $Mn_2Sn$ | 0.5969 | 0.2174 | 0.1667 | 0% |
| $Ge_6Mn_6Zr$ | 0.4232 | 0.2202 | 0.1659 | 17% |
| $Mn_3O_6Te$ | 0.9771 | 0.1946 | 0.1641 | 6% |
| $Fe_3Se_4$ | 0.4797 | 0.2201 | 0.1605 | 4% |
| $CoF_3Na$ | 1.4834 | 0.2374 | 0.1597 | 6% |
| $Br_2Co$ | 0.5756 | 0.2632 | 0.1590 | 0% |
| $CsF_3Fe$ | 1.0429 | 0.2472 | 0.1588 | 0% |
| $GeMn_2Se_4$ | 0.2634 | 0.1333 | 0.1580 | 0% |
| CrSe | 0.5635 | 0.2456 | 0.1580 | 0% |
| $Ge_6Mn_6Tm$ | 0.3237 | 0.1618 | 0.1580 | >20% |
| $DyMnO_3$ | 1.0120 | 0.1865 | 0.1577 | >20% |
| AsMnRu | 0.4595 | 0.2117 | 0.1577 | 17% |
| $CInMn_3$ | 0.6449 | 0.2223 | 0.1575 | 0% |
| GeMnNb | 0.2113 | 0.1232 | 0.1569 | >20% |
| $Ge_6Mn_6Sc$ | 0.5040 | 0.2274 | 0.1561 | >20% |
| $Fe_7Mo_6$ | 0.3430 | 0.1158 | 0.1552 | >20% |
| $Ge_2Mn_2Yb$ | 0.5877 | 0.2344 | 0.1548 | >20% |
| $Be_2Fe$ | 0.6474 | 0.2050 | 0.1537 | 19% |
| $AgMn_3N$ | 0.6798 | 0.2383 | 0.1521 | 0% |
| $Fe_{16}OY_6$ | 0.5822 | 0.1811 | 0.1513 | >20% |
| $LaMn_2Si_2$ | 0.5444 | 0.1818 | 0.1507 | 0% |
| LuMnSi | 0.5231 | 0.1949 | 0.1503 | >20% |
| CrFeP | 0.7514 | 0.2193 | 0.1495 | >20% |
| $CaCr_2O_4$ | 0.6587 | 0.1943 | 0.1489 | 19% |
| MnRbSb | 0.5655 | 0.1861 | 0.1466 | 0% |
| MnNi | 0.4128 | 0.2064 | 0.1459 | >20% |
| AsMnV | 0.5226 | 0.2086 | 0.1458 | 6% |
| $Cr_2NiS_4$ | 0.7661 | 0.1762 | 0.1458 | >20% |
| $CrSe_2$ | 0.5831 | 0.1765 | 0.1454 | 0% |
| $Mn_2Si_2Th$ | 0.4167 | 0.1978 | 0.1453 | >20% |
| AsCrNi | 0.3856 | 0.1844 | 0.1452 | 17% |
| $BaCr_2O_4$ | 0.5269 | 0.1807 | 0.1445 | 12% |
| $B_2Fe_2Lu$ | 0.4252 | 0.1718 | 0.1443 | >20% |
| MnTe | 0.7655 | 0.2436 | 0.1432 | 0% |
| $Mn_{12}Th$ | 0.3367 | 0.2034 | 0.1432 | >20% |
| $GaMn_3N$ | 0.5708 | 0.2169 | 0.1420 | 8% |
| $CGaMn_3$ | 0.4534 | 0.1770 | 0.1398 | 8% |
| CaMnSi | 0.3517 | 0.1250 | 0.1381 | 0% |
| $CaGe_2Mn_2$ | 0.5491 | 0.1332 | 0.1358 | 19% |
| $Cr_5P_3Zr$ | 0.2515 | 0.1070 | 0.1351 | >20% |
| $HoMnO_3$ | 0.5537 | 0.1734 | 0.1334 | >20% |
| $Mn_7NaO_{12}$ | 0.5089 | 0.2002 | 0.1330 | 18% |
| $Cl_2Fe$ | 0.6836 | 0.2327 | 0.1329 | 0% |
| $CrSb_2$ | 0.3239 | 0.1569 | 0.1310 | 0% |
| $BaFeO_3$ | 0.9237 | 0.1812 | 0.1308 | >20% |
| $CrLaO_3$ | 0.4936 | 0.1617 | 0.1306 | 17% |
| $Fe_2P$ | 0.3487 | 0.1903 | 0.1291 | >20% |
| $Ni_3O_6Te$ | 0.2655 | 0.0865 | 0.1277 | 11% |
| $FeI_2$ | 0.2982 | 0.1457 | 0.1275 | 0% |
| $Co_3Er$ | 0.6966 | 0.2234 | 0.1265 | >20% |
| $GeMn_2S_4$ | 0.9142 | 0.1918 | 0.1262 | >20% |
| $Co_5Dy$ | 0.6852 | 0.1996 | 0.1258 | >20% |
| $BaCoO_3$ | 1.0460 | 0.1876 | 0.1254 | >20% |
| $Fe_{16}Ho_6O$ | 0.4508 | 0.1561 | 0.1253 | >20% |
| $Cr_2O_4Sr$ | 0.4056 | 0.1062 | 0.1239 | >20% |
| $Cl_3FeK$ | 0.6086 | 0.2059 | 0.1229 | 0% |
| $CaMnO_3$ | 0.5507 | 0.1454 | 0.1221 | 12% |
| MnSe | 0.3324 | 0.1815 | 0.1220 | 0% |
| $Cr_2O_6Te$ | 0.4694 | 0.1481 | 0.1216 | 19% |
| GaMnPt | 0.5290 | 0.1917 | 0.1214 | 19% |
| $BaCrS_2$ | 0.7083 | 0.1753 | 0.1203 | 19% |
| $DyGe_6Mn_6$ | 0.2921 | 0.1138 | 0.1203 | >20% |
| $CeCo_3Pu_4$ | 0.3337 | 0.1565 | 0.1188 | >20% |
| $Co_2CrGa$ | 0.4882 | 0.1925 | 0.1183 | >20% |
| $MnTiO_3$ | 1.4424 | 0.1578 | 0.1181 | >20% |
| $MnSe_2$ | 0.3644 | 0.1349 | 0.1167 | 0% |
| $Mn_3NZn$ | 0.7558 | 0.2319 | 0.1160 | 8% |
| $Cr_2Te_3$ | 0.5523 | 0.1581 | 0.1152 | 0% |
| $B_4Co_{21}Nd_5$ | 0.5733 | 0.1611 | 0.1151 | >20% |
| $Br_2Fe$ | 0.5150 | 0.1919 | 0.1147 | 0% |
| $NiSmO_3$ | 0.4200 | 0.1839 | 0.1131 | 19% |
| KMnSb | 0.4899 | 0.1589 | 0.1122 | 0% |
| $HoMn_2Si_2$ | 0.3415 | 0.1202 | 0.1118 | >20% |
| $Cr_2P$ | 0.1033 | 0.0549 | 0.1086 | >20% |
| $MnYO_3$ | 0.5655 | 0.1720 | 0.1080 | >20% |
| $CrNdO_3$ | 0.4394 | 0.1399 | 0.1065 | 19% |
| CaMnSn | 0.3725 | 0.1355 | 0.1058 | 0% |
| $Cr_3Pt$ | 0.3489 | 0.1583 | 0.1052 | 8% |
| $FeNdSb_2$ | 0.4199 | 0.2064 | 0.1044 | 0% |
| $MnSb_2Sr_2$ | 0.4472 | 0.1219 | 0.1040 | 8% |
| BrCrO | 0.2449 | 0.0949 | 0.1037 | 0% |
| $BCeCo_4$ | 0.5128 | 0.1715 | 0.1031 | >20% |
| GeMnPd | 0.4283 | 0.1747 | 0.1023 | 8% |
| $FeH_2O_2$ | 0.2393 | 0.0864 | 0.1017 | 0% |
| $As_2Ni_3O_8$ | 0.1998 | 0.0696 | 0.1015 | 0% |
| GaNiU | 0.5570 | 0.1358 | 0.1005 | >20% |
| $CrLiO_2$ | 0.2467 | 0.0844 | 0.1004 | 0% |
| $CrHO_2$ | 0.2286 | 0.0637 | 0.1001 | 0% |
| $Bi_6Mn_3Sm_2$ | 0.3222 | 0.0814 | 0.0991 | 0% |
| $BeCr_2O_4$ | 0.2990 | 0.1333 | 0.0968 | 0% |
| $CN_2Ni$ | 0.3865 | 0.1575 | 0.0962 | >20% |
| $CrDyO_3$ | 0.3653 | 0.1249 | 0.0959 | 0% |
| $Mn_2O_4Ti$ | 0.5143 | 0.1347 | 0.0945 | >20% |
| $CrCuS_2$ | 0.2689 | 0.0980 | 0.0936 | 8% |
| $CoH_2O_2$ | 0.3269 | 0.1240 | 0.0925 | 0% |
| $CCoO_3$ | 0.3085 | 0.1193 | 0.0884 | 8% |
| $FeI_3Tl$ | 0.5697 | 0.1449 | 0.0839 | 0% |
| $FeHO_2$ | 0.2799 | 0.0954 | 0.0828 | 12% |
| $Bi_2CaMn$ | 0.3036 | 0.0977 | 0.0828 | 0% |
| $Nb_2Ni_4O_9$ | 0.3740 | 0.1129 | 0.0821 | >20% |
| $Mn_2Si_2Tb$ | 0.3202 | 0.1021 | 0.0818 | >20% |
| $CeCrS_3$ | 0.2937 | 0.0696 | 0.0794 | 0% |
| $Cr_2HoSi_2$ | 0.2877 | 0.0698 | 0.0786 | >20% |
| $LiMnO_2$ | 0.1803 | 0.0644 | 0.0781 | 0% |
| $AlB_2Fe_2$ | 0.2533 | 0.1140 | 0.0746 | >20% |
| $Fe_2ILa_2$ | 0.3521 | 0.1412 | 0.0738 | >20% |
| $FePrSb_3$ | 0.3023 | 0.0691 | 0.0716 | >20% |
| $MnSrO_3$ | 0.8484 | 0.1064 | 0.0695 | 12% |
| $AgCrO_2$ | 0.3090 | 0.1125 | 0.0694 | >20% |
| NiSbV | 0.2399 | 0.0822 | 0.0686 | >20% |
| NiO | 0.3596 | 0.1021 | 0.0681 | >20% |
| $AlB_2Mn_2$ | 0.1863 | 0.0660 | 0.0673 | >20% |
| $BiNiO_3$ | 0.4270 | 0.0987 | 0.0661 | >20% |
| $B_2FeMo_2$ | 0.2260 | 0.0756 | 0.0618 | >20% |
| $CeCrSe_3$ | 0.2060 | 0.0658 | 0.0602 | 12% |
| $ErFe_2Ge_2$ | 0.2946 | 0.1001 | 0.0600 | >20% |
| $FeSb_2$ | 0.2154 | 0.0991 | 0.0582 | >20% |
| GdMnSi | 0.3236 | 0.0922 | 0.0519 | >20% |
| $Au_{31}Mn_9$ | 0.2844 | 0.0789 | 0.0508 | >20% |
| $Cl_3KNi$ | 0.4912 | 0.0762 | 0.0433 | 8% |
| $CrFe_2Ga$ | 0.2225 | 0.0664 | 0.0383 | >20% |
| $CoSmO_3$ | 0.2343 | 0.0558 | 0.0295 | >20% |

Appendix B: Pressure list

ΔPressure: pressure difference between the ferromagnetic and paramagnetic phases.
Spin flip %: percentage of magnetic atoms that have changed sign from their initial setting.

| Compound | ΔPressure | Spin flip % |
|---|---|---|
| $BFe_{14}Ho_2$ | -71.6750 | >20% |
| $BFe_{14}Tb_2$ | -69.4300 | >20% |
| $Fe_{17}Pr_2$ | -64.4950 | 8% |
| $Fe_{17}Nd_2$ | -63.9750 | 12% |
| $BFe_{14}Gd_2$ | -63.1100 | >20% |
| $BFe_{14}Pr_2$ | -62.6600 | 5% |
| $Ce_2Fe_{17}$ | -58.6950 | 14% |
| $Fe_{17}Gd_2$ | -56.6950 | 12% |
| $Er_2Fe_{17}$ | -55.9700 | >20% |
| $BCe_2Fe_{14}$ | -53.6850 | 5% |
| MnO | -53.2300 | 6% |
| $Fe_3Pt$ | -51.4900 | >20% |
| $CFe_3$ | -50.7600 | 0% |
| $CuFeO_2$ | -48.8800 | >20% |
| $FeNaO_2$ | -48.5850 | 12% |
| $Fe_2Ti$ | -48.4050 | >20% |
| $AlFeO_3$ | -44.8500 | 12% |
| FeN | -41.4600 | >20% |
| $Mn_7NaO_{12}$ | -41.0450 | 18% |
| $Mn_2O_3$ | -40.3450 | 6% |
| $Co_5Dy$ | -39.7400 | >20% |
| $CrMn_2O_4$ | -39.4650 | 0% |
| CoFe | -38.9400 | >20% |
| $BFe_{14}Nd_2$ | -38.6150 | 5% |
| $Fe_3Pd$ | -37.5950 | 12% |
| $Fe_2O_3$ | -37.5750 | 19% |
| $Co_5Er$ | -37.2500 | >20% |
| $C_2Fe_5$ | -35.3800 | 20% |
| MnS | -34.6600 | 0% |
| $C_3Fe_{17}Tb$ | -34.5450 | 18% |
| $BFe_3O_5$ | -32.8800 | 8% |
| $CaMn_7O_{12}$ | -31.7750 | 14% |
| $Co_5Y$ | -30.4750 | >20% |
| $AgFeO_2$ | -29.1500 | 17% |
| $Fe_3N$ | -28.2350 | 8% |
| MnNi | -27.9850 | >20% |
| $Co_5Nd$ | -27.9800 | 5% |
| $Co_5Gd$ | -26.7600 | >20% |
| $AgMn_3N$ | -25.9050 | 0% |
| $Fe_3O_7P$ | -25.3400 | 6% |
| $Cr_2O_3$ | -25.1800 | 0% |
| $MnS_2$ | -24.2900 | 12% |
| $DyFe_5$ | -24.1000 | 20% |
| $MnSe$ | -23.2900 | 0% |
| $BMn_3O_5$ | -23.1900 | 17% |
| $Fe_{13}PbPr_6$ | -22.9850 | 15% |
| $C_2Fe_{11}Th$ | -22.7150 | 18% |
| $MnO_2$ | -22.4400 | 0% |
| $Fe_5Th$ | -22.3600 | 10% |
| $MnSe_2$ | -22.0050 | 0% |
| $Fe_5Nd$ | -21.4350 | 20% |
| $Fe_5Na_3O_9$ | -21.1450 | 10% |
| | | 0 |
| $Cl_3Mn_8O_{10}$ | -21.080 | 0% |
| $Fe_3Th$ | -20.9700 | >20% |
| $FeKO_2$ | -20.8600 | 6% |
| $Co_5Tb$ | -20.7750 | >20% |
| $DyMn_2O_5$ | -19.8450 | >20% |
| $LuMn_2O_5$ | -19.7900 | >20% |
| $NiO$ | -19.7300 | >20% |
| $BCo_{14}Nd_2$ | -19.4500 | >20% |
| $FeLuO_3$ | -19.3100 | >20% |
| $FeScO_3$ | -19.2100 | 8% |
| $As_2EuFe_2$ | -18.5450 | >20% |
| $Fe_6Ge_6Sc$ | -18.4100 | 17% |
| $Fe_2Ge_2Pr$ | -18.2650 | 0% |
| $GdMnO_3$ | -18.0300 | >20% |
| $GeMnNb$ | -18.0250 | >20% |
| $BiMn_2O_5$ | -17.9200 | 12% |
| $LaMnO_3$ | -17.9050 | 8% |
| $GeMn_2O_4$ | -17.8850 | 12% |
| $LiMn_2O_4$ | -17.8750 | >20% |
| $Fe_5Y$ | -17.8650 | >20% |
| $CeFe_5$ | -17.0600 | 15% |
| $CrSe$ | -16.8500 | 0% |
| $Fe_3Ga$ | -16.8200 | 0% |
| $GeMnY$ | -16.7500 | >20% |
| $Mn_2NaO_4$ | -16.5350 | >20% |
| $Fe_5Ho$ | -16.0450 | >20% |
| $Mn_3NRh$ | -15.9300 | 12% |
| $Mn_3NPt$ | -15.7650 | >20% |
| $BMn_2O_4$ | -15.5650 | 6% |
| $BErFe_4$ | -15.4650 | >20% |
| $Fe_4N$ | -15.4000 | 0% |
| GaHoMn | -15.2750 | >20% |
| $CeCo_3Pu_4$ | -15.2550 | >20% |
| $Mn_{12}Th$ | -15.2050 | >20% |
| $LiMnO_2$ | -15.2000 | 0% |
| $Co_5Pr$ | -15.0100 | 10% |
| $HoMnO_3$ | -14.3550 | >20% |
| $CrLiO_2$ | -14.1650 | 0% |
| $CaCr_2O_4$ | -14.1650 | 19% |
| $Fe_2NaO_3$ | -14.0550 | 12% |
| $MnSrO_3$ | -13.7650 | 12% |
| $Co_3Er$ | -13.1750 | >20% |
| FeGa | -13.1500 | 15% |
| $Cr_3Se_4$ | -13.1300 | 4% |
| $CrLaO_3$ | -13.0650 | 17% |
| $Fe_3Ho$ | -12.9900 | >20% |
| $MnYO_3$ | -12.7650 | >20% |
| $Mn_2O_5Yb$ | -12.6850 | >20% |
| $CGeMn_3$ | -12.6600 | 8% |
| $Cr_2O_6Re$ | -12.6250 | >20% |
| AsCrRh | -12.2700 | >20% |
| MnP | -12.0200 | 19% |
| $AuMn_2$ | -11.8900 | 9% |
| $MnYbO_3$ | -11.4950 | >20% |
| CrN | -11.3900 | 3% |
| $Fe_2S_4Si$ | -11.3750 | >20% |
| $MnNdO_3$ | -11.2300 | 12% |
| $Fe_4Ge_2Yb$ | -11.1550 | >20% |
| CoMnSi | -11.0200 | >20% |
| $MnTiO_3$ | -10.8750 | >20% |
| $Fe_2P$ | -10.8400 | >20% |
| $Ca_2Fe_7O_{11}$ | -10.5350 | 7% |
| $Fe_6Ge_6Hf$ | -10.5300 | >20% |
| $MnPrO_3$ | -10.3950 | 12% |
| $Ca_2Fe_2O_5$ | -10.3650 | >20% |
| $DyMnO_3$ | -10.3250 | >20% |
| $Cr_2O_4Sr$ | -10.3100 | >20% |
| $GdMn_2O_5$ | -10.2200 | >20% |
| $Fe_5Gd$ | -10.0750 | >20% |
| $Fe_3Ge_2O_8$ | -10.0550 | 0% |
| $BaMnO_3$ | -10.0400 | 0% |
| $EuMnO_3$ | -9.9650 | >20% |
| $CrO_2$ | -9.8850 | 12% |
| $Be_2Fe$ | -9.8500 | 19% |
| $B_2Fe_2Lu$ | -9.5800 | >20% |
| $Fe_2MgO_4$ | -9.4300 | >20% |
| $AgCrO_2$ | -9.3200 | >20% |
| $Fe_7Mo_6$ | -9.2650 | >20% |
| $BaFe_2Se_3$ | -9.1350 | 0% |
| $CaMnO_3$ | -9.1300 | 12% |
| $B_4Co_{21}Nd_5$ | -9.1250 | >20% |
| $AlB_2Fe_2$ | -9.0800 | >20% |
| $Mn_2O_5Sr_2$ | -8.8300 | >20% |
| $BeCr_2O_4$ | -8.7800 | 0% |
| $BCeCo_4$ | -8.7200 | >20% |
| $MnTbO_3$ | -8.4150 | >20% |
| $EuMn_2Sb_2$ | -8.2250 | 6% |
| $MnSmO_3$ | -8.2200 | >20% |
| $BaCrS_2$ | -8.1350 | 19% |
| $C_3Fe_{17}Th_2$ | -7.9950 | >20% |
| $Ge_6Mn_6Zr$ | -7.8250 | 17% |
| $BaCr_2O_4$ | -7.8200 | 12% |
| $CrHO_2$ | -7.7900 | 0% |
| $BiMnO_3$ | -7.5550 | >20% |
| ClFeO | -7.4050 | 6% |
| $CN_2Ni$ | -7.2000 | >20% |
| AsCrNi | -6.8600 | 17% |
| FeSn | -6.7500 | 0% |
| $CFe_2$ | -6.7250 | 6% |
| $FeO_2$ | -6.5300 | 6% |
| $CrCuS_2$ | -6.4600 | 8% |
| $Fe_3Sn$ | -6.4250 | 8% |
| $Fe_{16}OY_6$ | -6.4150 | >20% |
| $CrNdO_3$ | -6.1750 | 19% |
| $CoSmO_3$ | -6.0400 | >20% |
| $CsFe_2Se_2$ | -5.9600 | 3% |
| FeS | -5.6550 | 6% |
| GaMn | -5.5600 | 3% |
| $Fe_2KSe_2$ | -5.5450 | 9% |
| $CrDyO_3$ | -5.4000 | 0% |
| $CuFe_2Ge_2$ | -5.3850 | 19% |
| $Fe_6Ge_4Li$ | -5.1300 | 17% |
| $BFe_2O_4$ | -4.9100 | 6% |
| $Cr_2Te_3$ | -4.8550 | 0% |
| $Mn_2Sb_2Yb$ | -4.8500 | 0% |
| GeMnSr | -4.6700 | 6% |
| $Fe_{16}Ho_6O$ | -4.5650 | >20% |
| $CuFeSe_2$ | -4.3600 | >20% |
| $CrSe_2$ | -4.3450 | 0% |
| BMn | -4.1300 | 19% |
| $BaFe_4O_7$ | -4.1100 | 19% |
| MnRbSb | -4.0700 | 0% |
| AsMnRu | -3.9700 | 17% |
| KMnSb | -3.7150 | 0% |
| $Co_2CrGa$ | -3.6550 | >20% |
| $Fe_2Th$ | -3.6450 | >20% |
| $Mn_2O_4Ti$ | -3.6150 | >20% |
| $CrSb_2$ | -3.4850 | 0% |
| $NiSmO_3$ | -3.3400 | 19% |
| $Mn_3O_6Te$ | -3.3100 | 6% |
| $MnSb_2Sr_2$ | -3.2400 | 8% |
| $Cr_2P$ | -3.1350 | >20% |
| $CeCrS_3$ | -2.9150 | 0% |
| $BiNiO_3$ | -2.9000 | >20% |
| $Nb_2Ni_4O_9$ | -2.8100 | >20% |
| BrCrO | -2.7600 | 0% |
| $Ca_2Co_2O_5$ | -2.6550 | 19% |
| $Fe_3Te_3Tl$ | -2.4900 | 8% |
| $GeMn_2Se_4$ | -2.4100 | 0% |
| $Au_{31}Mn_9$ | -1.9950 | >20% |
| GaMnPt | -1.8200 | 19% |
| $CrFe_2Ga$ | -1.7200 | >20% |
| FeSb | -1.7100 | 12% |
| $FeNdSb_2$ | -1.4600 | 0% |
| $Co_2FeIn$ | -1.3350 | >20% |
| BaGeMn | -0.9900 | 0% |
| $Ni_3O_6Te$ | -0.9250 | 11% |
| $Bi_6Mn_3Sm_2$ | -0.8950 | 0% |
| $Fe_2Rb_3Se_4$ | -0.7700 | 0% |
| AsMnV | -0.6600 | 6% |
| $BaFeO_3$ | -0.6350 | >20% |
| $CeCrSe_3$ | -0.6200 | 12% |
| $As_2Ni_3O_8$ | -0.5800 | 0% |
| CrS | -0.4500 | 6% |
| $GeMn_2S_4$ | -0.4050 | >20% |
| GeHoMn | -0.3950 | >20% |
| CaMnSn | -0.2450 | 0% |
| $B_2FeMo_2$ | -0.1050 | >20% |
| $Ge_6Mn_6Sc$ | -0.1050 | >20% |
| $BaCoO_3$ | 0.1350 | >20% |
| MnTe | 0.1350 | 0% |
| $Cr_2NiS_4$ | 0.2800 | >20% |
| $AsFe_2$ | 0.3600 | 19% |
| $Fe_2ILa_2$ | 0.4300 | >20% |
| $CsF_3Fe$ | 0.4900 | 0% |
| $ErFeO_3$ | 0.5300 | >20% |
| $As_3CaFe_5$ | 0.6400 | 10% |
| $Mn_5Si_2$ | 1.0950 | >20% |
| $Fe_5Si_3$ | 1.1800 | 10% |
| $Cr_2O_6Te$ | 1.2150 | 19% |
| GdMnSi | 1.4150 | >20% |
| $FeTmO_3$ | 1.5550 | >20% |
| GaNiU | 1.7600 | >20% |
| $Ge_6Mn_6Tm$ | 1.7650 | >20% |
| $Mn_3NNi$ | 1.8500 | 12% |
| $BaFe_{12}O_{19}$ | 1.9100 | 8% |
| $DyGe_6Mn_6$ | 1.9700 | >20% |
| $CInMn_3$ | 2.0100 | 0% |
| GeMnTm | 2.0300 | >20% |
| MnSiTb | 2.2200 | >20% |
| $Cr_5S_6$ | 2.3200 | 0% |
| NiSbV | 2.3200 | >20% |
| CaMnSi | 2.6100 | 0% |
| $Bi_2CaMn$ | 2.8100 | 0% |
| $Cr_2S_3$ | 2.8300 | 12% |
| $FeI_3Tl$ | 2.8650 | 0% |
| $Fe_2MnSi$ | 3.4450 | >20% |
| CoO | 3.5150 | 6% |
| $Mn_3NPd$ | 3.6150 | 17% |
| $ErFe_2Ge_2$ | 3.6300 | >20% |
| $Mn_2Si_2Th$ | 3.6750 | >20% |
| $Mn_2Si_2U$ | 3.9900 | >20% |
| $Fe_3Se_4$ | 4.1400 | 4% |
| $Fe_3S_4Tl_2$ | 4.1500 | >20% |
| $Fe_2S_4Tl_3$ | 4.3850 | 19% |
| $Fe_2Na_3S_4$ | 4.9200 | 19% |
| $Mn_2Si_2Tb$ | 4.9800 | >20% |
| $Mn_2NpSi_2$ | 5.0350 | 4% |
| $Cl_3KNi$ | 5.1400 | 8% |
| $FePrSb_3$ | 5.2750 | >20% |
| $Fe_4O_5$ | 5.4650 | 12% |
| $Mn_2P$ | 5.8600 | 17% |
| $HoMn_2Si_2$ | 5.9950 | >20% |
| CrSb | 6.0000 | 0% |
| $AlB_2Mn_2$ | 6.0500 | >20% |
| $Fe_2Sn$ | 6.0600 | 0% |
| AlGeMn | 6.1350 | 6% |
| DyGeMn | 6.3050 | >20% |
| $FeTiO_3$ | 6.4650 | >20% |
| $Ge_2Mn_2Yb$ | 6.6600 | >20% |
| $Mn_2Si_2Y$ | 6.8250 | >20% |
| $FeNdO_3$ | 7.6800 | >20% |
| $Fe_2N$ | 7.9250 | 17% |
| GeMnYb | 8.1900 | 8% |
| $Cr_5P_3Zr$ | 8.2650 | >20% |
| $CaGe_2Mn_2$ | 8.4900 | 19% |
| $Fe_6Ge_6Li$ | 8.6200 | 6% |
| $EuMn_2P_2$ | 8.8550 | 6% |
| LuMnSi | 9.0100 | >20% |
| BFe | 9.2200 | >20% |
| FeSe | 9.3000 | 6% |
| $FeSb_2$ | 9.3050 | >20% |
| $Mn_2O_3Ta$ | 9.6000 | 12% |
| GeMnPd | 9.6650 | 8% |
| $Cr_2HoSi_2$ | 9.9000 | >20% |
| AsMnNa | 9.9900 | 0% |
| $LaMn_2Si_2$ | 10.0650 | 0% |
| MnSiYb | 10.3600 | 0% |
| $Cr_2Si_2U$ | 10.8950 | >20% |
| $Cl_2Fe$ | 11.3100 | 0% |
| $FeHO_2$ | 11.6250 | 12% |
| $Mn_2Sn$ | 13.1400 | 0% |
| $Fe_6Ge_6Mg$ | 13.4300 | 0% |
| $FeLaO_3$ | 13.6650 | 12% |
| AsMnRh | 14.0200 | >20% |
| $Mn_6Sn_6Tb$ | 14.0250 | >20% |
| $FeYO_3$ | 14.4300 | >20% |
| $Mn_2Sb$ | 14.8500 | 0% |
| $Ge_2Mn_2U$ | 15.2050 | 0% |
| CrFeP | 15.6600 | >20% |
| $BiFeO_3$ | 16.2250 | 6% |
| $FeI_2$ | 16.3350 | 0% |
| $CoF_3$ | 17.5550 | >20% |
| $BCo_4Y$ | 18.8650 | >20% |
| $Fe_7O_9$ | 19.1400 | 14% |
| $MnSnO_3$ | 20.1300 | 17% |
| GdGeMn | 21.8550 | >20% |
| $As_2Mn_2Zn$ | 22.0650 | 0% |
| $Cl_3FeK$ | 22.2850 | 0% |
| $As_2EuMn_2$ | 23.2150 | 0% |
| $AsMn_2$ | 23.2600 | 0% |
| $CGaMn_3$ | 24.9000 | 8% |
| MnN | 25.4450 | 0% |
| $Mn_3NZn$ | 25.6750 | 8% |
| $Cr_3Pt$ | 30.8400 | 8% |
| $F_6Fe_2Li$ | 31.2750 | 6% |
| $FeH_2O_2$ | 34.6000 | 0% |
| $Br_2Fe$ | 34.6200 | 0% |
| $Mn_3Sn$ | 37.6000 | 0% |
| $F_6Fe_2Rb$ | 37.8800 | 19% |
| $Fe_2O_4Si$ | 38.5350 | 12% |
| $CCoO_3$ | 40.2000 | 8% |
| $As_2Mn_2Yb$ | 41.3100 | 0% |
| $GeMn_3$ | 45.6950 | 0% |
| AsLiMn | 47.9850 | 0% |
| $GaMn_3N$ | 48.5650 | 8% |
| $Mn_3Sb$ | 52.1200 | 0% |
| $FeSiO_3$ | 55.5500 | 19% |
| $Mn_2O_4Si$ | 58.0300 | 19% |
| $CoTiO_3$ | 61.9250 | >20% |
| $Fe_3O_4$ | 65.8850 | 8% |
| $CoH_2O_2$ | 66.6650 | 0% |
| $As_3Co_5O_{12}$ | 69.6850 | 0% |
| $CoF_3K$ | 85.4800 | 12% |
| $Br_2Co$ | 97.3800 | 0% |
| $CoF_3Na$ | 141.9150 | 6% |
| $Mn_3Pt$ | 148.0050 | >20 |

## Appendix C: Top candidates by type

Max Δforce: maximum force difference between different DLM configurations.
Avg. Δforce: average of the force difference between different DLM configurations
Avg. force: Average force acting on each atom.
Spin flip %: percentage of magnetic atoms that have changed sign from their initial setting.

### Oxides

| Compound | Max Δforce. | Avg. Δforce. | Avg. force. | Spin flip % |
|---|---|---|---|---|
| $Fe_3O_4$ | 1.5156 | 0.6936 | 0.5700 | 8% |
| $Fe_2O_3$ | 1.4123 | 0.6477 | 0.5270 | 19% |
| $BaFe_4O_7$ | 1.0437 | 0.4012 | 0.4673 | 19% |
| $CrMn_2O_4$ | 1.2232 | 0.4193 | 0.4660 | 0% |
| $Fe_4O_5$ | 0.9972 | 0.4331 | 0.4645 | 12% |

### Nitrides

| Compound | Max Δforce. | Avg. Δforce. | Avg. force. | Spin flip % |
|---|---|---|---|---|
| $Fe_4N$ | 0.6483 | 0.4064 | 0.5128 | 0% |
| FeN | 0.8733 | 0.5797 | 0.4440 | >20% |
| CrN | 1.4970 | 0.6460 | 0.4257 | 3% |
| $Mn_3NRh$ | 1.2630 | 0.5677 | 0.4127 | 12% |
| MnN | 1.1663 | 0.5621 | 0.3765 | 0% |

### Halogens

| Compound | Max Δforce. | Avg. Δforce. | Avg. force. | Spin flip % |
|---|---|---|---|---|
| ClFeO | 0.8810 | 0.2402 | 0.2904 | 6% |
| $CoF_3$ | 1.8717 | 0.4938 | 0.2841 | >20% |
| $Cl_3Mn_8O_{10}$ | 0.6370 | 0.2455 | 0.1850 | 0% |
| $CoF_3K$ | 0.8017 | 0.2101 | 0.1676 | 12% |
| $CoF_3Na$ | 1.4834 | 0.2374 | 0.1597 | 6% |

### Intermetallics

| Compound | Max Δforce. | Avg. Δforce. | Avg. force. | Spin flip % |
|---|---|---|---|---|
| $Er_2Fe_{17}$ | 1.0150 | 0.4182 | 0.3897 | >20% |
| $Fe_2Ge_2Pr$ | 0.8272 | 0.2356 | 0.3642 | 0% |
| $Fe_{17}Nd_2$ | 0.7767 | 0.3241 | 0.3495 | 12% |
| $Fe_{17}Pr_2$ | 0.9366 | 0.3364 | 0.3402 | 8% |
| $Fe_3Sn$ | 0.8956 | 0.4349 | 0.3276 | 0% |

### Chalcogenides

| Compound | Max Δforce. | Avg. Δforce. | Avg. force. | Spin flip % |
|---|---|---|---|---|
| $CsFe_2Se_2$ | 1.1018 | 0.3221 | 0.3729 | 3% |
| $Cr_2S_3$ | 1.4882 | 0.3358 | 0.3237 | 12% |
| $BaFe_2Se_3$ | 0.4277 | 0.2085 | 0.3225 | 0% |
| $Fe_2KSe_2$ | 1.2490 | 0.4189 | 0.3185 | 9% |
| FeS | 0.8795 | 0.3006 | 0.3128 | >20% |

### Pictinides

| Compound | Max Δforce. | Avg. Δforce. | Avg. force. | Spin flip |
|---|---|---|---|---|
| $As_2EuFe_2$ | 0.8354 | 0.3146 | 0.3370 | >20% |
| $BiFeO_3$ | 1.0374 | 0.3306 | 0.3282 | 6% |
| $As_3CaFe_5$ | 0.7692 | 0.3204 | 0.3266 | 10% |
| $Mn_2P$ | 0.6437 | 0.2416 | 0.3230 | >20% |
| AsLiMn | 0.6638 | 0.2389 | 0.3223 | 0% |